\documentclass[twocolumn,twocolappendix]{aastex7}

\usepackage{amsmath}
\usepackage{txfonts}
\usepackage{threeparttable}
\hypersetup{allcolors=blue}
\usepackage{graphicx}
\usepackage{txfonts}
\newcommand\iona[2]{#1$\;${\scshape{#2}}}

\newcommand{\focc}{$f_\mathrm{occ}$}
\newcommand{\mstar}{$M_\star$}
\newcommand{\msun}{$M_{\odot}$}
\newcommand{\mbh}{$M_{\rm BH}$}
\newcommand{\gtsima}{$\; \buildrel > \over \sim \;$}
\newcommand{\ltsima}{$\; \buildrel < \over \sim \;$}
\newcommand{\lsim}{\lower.5ex\hbox{\ltsima}}
\newcommand{\simgt}{\lower.5ex\hbox{\gtsima}}
\newcommand{\simlt}{\lower.5ex\hbox{\ltsima}}

\newcommand{\lx}{$L_{\rm X}$}

\begin{document}
\title{Central Massive Black Holes Are Not Ubiquitous in Local Low-Mass Galaxies}

\author[0000-0002-4436-6923]{Fan Zou}
\affiliation{Department of Astronomy, University of Michigan, 1085 S University, Ann Arbor, MI 48109, USA}
\email[show]{fanzou01@gmail.com}

\author[0000-0001-5802-6041]{Elena Gallo}
\affiliation{Department of Astronomy, University of Michigan, 1085 S University, Ann Arbor, MI 48109, USA}
\email{egallo@umich.edu}

\author[0000-0003-0248-5470]{Anil C. Seth}
\affiliation{Department of Physics and Astronomy, University of Utah}
\email{aseth@astro.utah.edu}

\author[0000-0002-2397-206X]{Edmund Hodges-Kluck}
\affiliation{X-ray Astrophysics Laboratory, NASA/GSFC, Greenbelt, MD 20771, USA}
\email{Edmund.hodges-kluck@nasa.gov }

\author[0009-0004-9457-2495]{David Ohlson}
\affiliation{Department of Physics and Astronomy, University of Utah}
\email{david.ohlson@utah.edu}

\author[0000-0002-8460-0390]{Tommaso Treu}
\affiliation{Physics and Astronomy Department, University of California, Los Angeles, CA 90095}
\email{tt@astro.ucla.edu}

\author[0000-0003-4703-7276]{Vivienne F. Baldassare}
\affiliation{Department of Physics and Astronomy, Washington State University, Pullman, WA 99164, USA}
\email{vivienne.baldassare@wsu.edu}

\author[0000-0002-0167-2453]{W.N. Brandt}
\affiliation{Department of Astronomy and Astrophysics, 525 Davey Lab, The Pennsylvania State University, University Park, PA 16802, USA}
\affiliation{Institute for Gravitation and the Cosmos, The Pennsylvania State University, University Park, PA 16802, USA}
\affiliation{Department of Physics, 104 Davey Laboratory, The Pennsylvania State University, University Park, PA 16802, USA}
\email{wnbrandt@gmail.com}

\author[0000-0002-5612-3427]{Jenny E. Greene}
\affiliation{Department of Astrophysical Sciences, Princeton University, 4 Ivy Lane, Princeton, NJ 08544, USA}
\email{jgreene@astro.princeton.edu }

\author[0000-0002-6336-3293]{Piero Madau}
\affiliation{Department of Astronomy \& Astrophysics, University of California, 1156 High Street, Santa Cruz, CA 95064, USA}
\affiliation{Dipartimento di Fisica ``G. Occhialini'', Universit{\'a} degli Studi di Milano-Bicocca, Piazza della Scienza 3, I-20126 Milano, Italy}
\email{pmadau@ucsc.edu}

\author[0000-0002-5678-1008]{Dieu D. Nguyen}
\affiliation{Department of Astronomy, University of Michigan, 1085 S University, Ann Arbor, MI 48109, USA}
\email{dieun@umich.edu}

\author[0000-0002-7092-0326]{Richard M. Plotkin}
\affiliation{Physics Department, University of Nevada, Reno, 1664 N. Virginia St, Reno, NV, 89557, USA.}
\affiliation{Nevada Center for Astrophysics, University of Nevada, Las Vegas, NV 89154, USA}
\email{rplotkin@unr.edu}

\author[0000-0001-7158-614X]{Amy E. Reines}
\affiliation{eXtreme Gravity Institute, Department of Physics, Montana State University, Bozeman, MT 59717, USA}
\email{amy.reines@montana.edu}

\author[0000-0003-4961-1606]{Alberto Sesana}
\affiliation{Dipartimento di Fisica ``G. Occhialini'', Universit{\'a} degli Studi di Milano-Bicocca, Piazza della Scienza 3, I-20126 Milano, Italy}
\affiliation{INFN, Sezione di Milano-Bicocca, Piazza della Scienza 3, I-20126 Milano, Italy}
\affiliation{INAF - Osservatorio Astronomico di Brera, via Brera 20, I-20121 Milano, Italy}
\email{alberto.sesana@unimib.it}

\author[0000-0002-8055-5465]{Jong-Hak Woo}
\affiliation{Astronomy Program, Department of Physics and Astronomy, Seoul National University, Seoul, 08826, Republic of Korea}
\email{jhwoo@snu.ac.kr}

\author[0000-0001-7349-4695]{Jianfeng Wu}
\affiliation{Department of Astronomy, Xiamen University, Xiamen, Fujian 361005, People's Republic of China}
\email{wujianfeng@xmu.edu.cn}

\defcitealias{Burke25}{B25}
\defcitealias{Miller15}{M15}

\begin{abstract}
The black-hole occupation fraction ($f_\mathrm{occ}$) defines the fraction of galaxies that harbor central massive black holes (MBHs), irrespective of their accretion activity level. While it is widely accepted that $f_\mathrm{occ}$ is nearly 100\% in local massive galaxies with stellar masses $M_\star \gtrsim 10^{10}~M_\odot$, it is not yet clear whether MBHs are ubiquitous in less-massive galaxies. 
In this work, we present new constraints on $f_\mathrm{occ}$ based on over 20 years of Chandra imaging data for 1606 galaxies within 50 Mpc. 
We employ a Bayesian model to simultaneously constrain $f_\mathrm{occ}$ and the specific accretion-rate distribution function, $p(\lambda)$, where the specific accretion rate is defined as $\lambda=L_\mathrm{X}/M_\star$, and $L_\mathrm{X}$ is the MBH accretion luminosity in the 2-10 keV range. Notably, we find that $p(\lambda)$ peaks around $10^{28}~\mathrm{erg~s^{-1}}~M_\odot^{-1}$; above this value, $p(\lambda)$ decreases with increasing $\lambda$, following a power-law that smoothly connects with the probability distribution of bona-fide active galactic nuclei.
We also find that the occupation fraction decreases dramatically with decreasing $M_\star$: in high mass galaxies ($M_\star \approx 10^{11-12}M_\odot$), the occupation fraction is $>93\%$ (a $2\sigma$ lower limit), and then declines to $66_{-7}^{+8}\%$ ($1\sigma$ errors) between $M_\star\approx10^{9-10}M_\odot$, and to $33_{-9}^{+13}\%$ in the dwarf galaxy regime between $M_\star\approx10^{8-9}~M_\odot$. Our results have significant implications for the normalization of the MBH mass function over the mass range most relevant for tidal disruption events, extreme mass ratio inspirals, and MBH merger rates that upcoming facilities are poised to explore.
%
\end{abstract}

\section{Introduction}
\label{sec: intro}
Massive black holes (MBHs) residing at the centers of nearby galaxies typically accrete material at rates far below their Eddington limits (e.g., \citealt{Ho09, Kormendy13}), making them comparatively difficult to identify using the diagnostic techniques conventionally applied to active galactic nuclei (AGN). Dynamical methods--tracing the motion of stars and gas in the vicinity of the MBH--bypass this limitation, indicating that nearly all Local Volume galaxies as large as the Milky Way, or larger, host an MBH. One of the most thoroughly studied examples is the Galactic Center MBH, Sgr A*, which has a dynamical black hole mass of $M_\mathrm{BH} \approx 4 \times 10^6~M_\odot$ \citep{Ghez2008, Gillessen2009} and an Eddington ratio close to $10^{-9}$ \citep{EHT22}.

It remains unclear whether central MBHs are ubiquitous in lower-mass galaxies, particularly dwarf galaxies with stellar masses of $M_\star\lesssim10^{9.5}~M_\odot$ (\citealt{Greene20, Reines22} and references therein). This stellar mass range corresponds to black hole masses of $M_\mathrm{BH}\lesssim10^6~M_\odot$, based on the local $M_\mathrm{BH}{:}M_\star$ scaling relation established for massive galaxies (e.g., \citealt{Reines15}), albeit the scaling relation in the dwarf regime is poorly constrained (e.g., \citealt{Greene20}). Due to the small sphere of influence, dynamical evidence for MBHs can only be acquired in a handful of such dwarf galaxies \citep{Nguyen18, Nguyen19, Greene20}.\\

Quantifying the active fraction ($f_a$; e.g., \citealt{Baldassare20, Zou23, Ohlson24, Pucha25}) and, more importantly, the true black-hole occupation fraction (\focc) as a function of \mstar\ is crucial to several high-impact astrophysical problems. For example, the occupation fraction in dwarf galaxies is thought to be highly sensitive to the black hole seeding mechanism at high redshift \citep{Volonteri12, Natarajan14, Agarwal16, Valiante16, Ricarte18b, Woods19, Inayoshi20}. Models suggest that, by $z=0$, virtually all dwarf galaxies will contain a MBH if stellar-sized, Pop III remnants ($M_{\bullet}\approx 10^2$ \msun; e.g., \citealt{Fryer01}) provide the {\it dominant} seeding mode, as opposed to global gas collapse on galaxy-size scales ($M_{\bullet}\approx 10^{5-6}$ \msun; e.g., \citealt{Habouzit16}), which leads to extremely low occupation fractions in dwarfs; if seeding occurs preferentially via gravitational runaway in dense stellar clusters ($M_{\bullet}\approx 10^{3-4}$ \msun; e.g., \citealt{Sakurai17}), then about 50\% of local dwarfs can be expected to host a MBH. The expected rates of extreme mass ratio inspirals \citep{Babak17} and tidal disruption events \citep{Stone16} that will be detectable by the Vera C. Rubin Observatory (Rubin; \citealt{Ivezic19}) and, later, the Laser Interferometer Space Antenna (LISA; \citealt{Amaro-Seoane17}) are highly sensitive to the occupation fraction in dwarf galaxies.  Additionally, {\it if} the occupation fraction of dwarf galaxies were close to 100\%, then their MBHs (rather than supernovae) could be entirely responsible for quenching star formation in these systems \citep{Dashyan18, Penny18, Dickey19}. Pursuing a local approach to the MBH seeding conundrum is particularly compelling because it is orthogonal and complementary to ongoing searches at very high redshifts with the James Webb Space Telescope. These observations appear to be more easily explained with heavy seeding (e.g., \citealt{Bogdan24, Greene24, Maiolino24, Natarajan24}). \par 

Hard \mbox{X-rays} ($\simgt 2$ keV) are arguably the best electromagnetic signpost of MBH activity within the Local Volume because the \mbox{X-ray} sky is dark, and even quiescent MBHs emit non-thermal \mbox{X-rays} due to persistent, low Eddington-ratio accretion \citep{Pellegrini10, Miller12, She17, Bi20}; furthermore, hard \mbox{X-rays} are highly penetrating and significantly less diluted by galaxy starlight (e.g., \citealt{Brandt15, Brandt22}) compared to both soft \mbox{X-rays} as well as optical photons. AGN in dwarf galaxies can be identified through multiple diagnostics (e.g., \citealt{Reines13, Sartori15, Baldassare20, Birchall20, Birchall22, Burke22, Reines22, Wasleske24}) and to cosmological distances \citep{Pardo16, Mezcua18, Mezcua23, Zou23}, even though they account for only a small fraction of the underlying population. These studies generally find that $f_a$ is a few percent or less. Such low fraction is in rough agreement with theoretical expectations \citep{Pacucci21}.
\par
Robustly constraining \focc\ based on $f_a$ requires leveraging a large, unbiased galaxy sample that spans as wide as possible a range of stellar masses down to the lowest possible Eddington ratios. 
However, there is an inherent limit to how faint or distant we can probe: due to contamination from \mbox{X-ray} binaries (XRBs) to the galaxy nuclear \mbox{X-ray} signal, the chosen \mbox{X-ray} instrument's Point Spread Function (PSF) effectively limits the distance within which this experiment can be efficiently carried out. In practice, the known scaling of XRB luminosity with stellar mass and star formation rate \citep{Lehmer19} sets a resolution- and distance-dependent limiting \mbox{X-ray} sensitivity, below which XRBs dominate the signal \citep{Hodges-Kluck20}. This limit is highly sensitive to spatial resolution, making the Chandra \mbox{X-ray} Observatory the only currently operating facility capable of making a constraining measurement of \focc\ in the local universe.\footnote{If selected, the Advanced X-ray Imaging Satellite (AXIS; \citealt{Reynolds23}) is expected to offer higher throughput at arcsecond-level resolution, thereby enabling even more effective \mbox{X-ray} based constraints \citep{Gallo23}.}
\par
Based on on-axis Chandra Advanced CCD Imaging Spectrometer (ACIS) observations of 326 nearby early-type galaxies, \cite{Gallo19} employ Bayesian inference to constrain the MBH occupation fraction as a function of \mstar, finding $f_\mathrm{occ}>47\%$ for host stellar masses in the range $10^{9-10}$ \msun\ (68\% C.L.), with no upper bound. Despite the improvement compared to the initial sample of approximately 200 early types considered by \citet[hereafter \citetalias{Miller15}]{Miller15}, \cite{Gallo19} have little or no constraining power in the bona-fide dwarf galaxy regime.\\

In this work, we expand on the aforementioned results by (i) leveraging the full Chandra Source Catalog\footnote{\url{https://cxc.cfa.harvard.edu/csc2.1/}} Release 2.1  (hereafter CSC 2.1; \citealt{Evans24}) and considering a total of about 1,600 target galaxies within 50 Mpc, and; (ii) foregoing the assumption of a log-linear relation with normal scatter between the MBH \mbox{X-ray} luminosity and the host stellar mass. \citet[hereafter \citetalias{Burke25}]{Burke25} recently derived $f_\mathrm{occ}$ based on similar underlying data, and we will discuss our significant differences in Section~\ref{subsec: comp_B25}.
\section{Sample and Data Description }
\label{sec: data}

\begin{figure}
\includegraphics[width=\hsize]{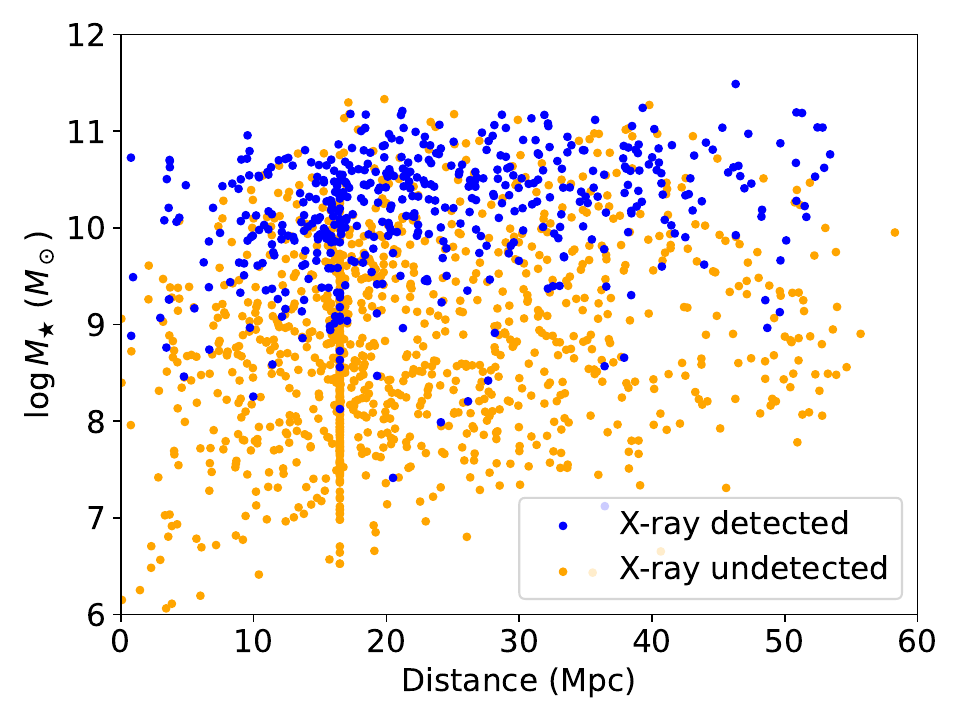}
\caption{$M_\star$ vs. distance for the target sample of 1,606 galaxies. Blue and orange points represent \mbox{X-ray}-detected and undetected galaxies, respectively. There are points visually ``piling up'' at a distance of 16.5~Mpc because they are from the Virgo Cluster (see Section~3.3 in \citealt{Ohlson24}).}
\label{fig: MstarDist}
\end{figure}

Our parent sample is drawn from the 50~Mpc Galaxy Catalog\footnote{\url{https://github.com/davidohlson/50MGC}} (50MGC; \citealt{Ohlson24}), which contains consistent and homogenized stellar mass, distance, and morphological type measurements for 15,424 galaxies within $\approx50$~Mpc. The 50MGC stellar masses are estimated through color-dependent mass-to-light ratios and are in good consistency with other works (see Section~4 of \citealt{Ohlson24}). The volume completeness is $>50\%$ above $10^9~M_\odot$ and $>20\%$ above $10^8~M_\odot$. To search for nuclear \mbox{X-ray} sources, we take advantage of the  CSC 2.1, which includes measured properties for 407,806 unique (compact and extended) \mbox{X-ray} sources detected in ACIS or High Resolution Camera (HRC) observations that were released publicly prior to the end of 2021. We only consider \mbox{X-ray} sources covered by ACIS because of the low energy resolution of HRC. This yields 1,606 unique galaxies with $M_\star$ measurements in the 50MGC within the ACIS footprint, which are shown in the plane of $M_\star$ vs. distance in Figure~\ref{fig: MstarDist}. Since only 10\% of the 50MGC sources are covered by ACIS, complex selection effects may arise. We address this issue in Appendix~\ref{append: selection}, and argue that the selection effects associated with intentional Chandra observations of MBHs are likely smaller than the statistical uncertainties.
\par

The CSC 2.1 Optical/IR Counterparts Catalog\footnote{\url{https://cxc.cfa.harvard.edu/csc/csc_crossmatches.html}} further provides the CSC 2.1 counterparts in Gaia DR3, Legacy Survey DR10, PanSTARRS-1 and 2MASS. Since these optical/IR positions are generally more accurate than \mbox{X-ray} ones, we start by matching the nominal positions of the 50MGC galaxies with the CSC 2.1 ACIS counterpart catalog using a radius of $1.5''$. This yields 477 \mbox{X-rays} sources. \citet{Ohlson24} already conducted careful visual inspections (including nuclear coordinates) and removed contaminants when constructing the 50MGC.
We also visually inspect the optical and \mbox{X-ray} images of our sources and find no apparent issues.
%

We use Equation 11 from \citet{Evans24} to determine in which energy bands each source was detected. For inferring \mbox{X-ray} photometry, we select the detected band based on the following priority: hard band ($2-7$~keV; 71\% of the sources), broad band ($0.5-7$~keV; 29\%), medium band ($1.5-2$~keV; one source only), soft band ($0.5-1.2$~keV; none), and ultra-soft band ($0.2-0.5$~keV; none). We prioritize higher energy bands to minimize absorption effects and reduce contamination from extended hot gas. We further exclude all the extended sources identified by the CSC's algorithm {\texttt{extent\_code}}. This yields a total of 388 central, point-like \mbox{X-ray} nuclei. For those, we estimate the corresponding $2-10$ keV flux ($f_\mathrm{X}$) using the aperture photometry within the 90\% enclosed-energy radius (i.e., \texttt{flux\_aper90} in the CSC) and assuming a photon index of 1.8 to convert the photometry in different energy bands to a common $2-10$~keV range. The uncertainty from the assumed photon index would effectively be absorbed into the width of the specific accretion-rate distribution function later on, which is measured to be $\sigma=1$~dex (see Section~\ref{sec: method} for more details). However, varying the photon index between 1.4 and 2.2 could cause an at-most 0.15~dex difference, which is much smaller than $\sigma=1$~dex and thus has minimal impacts. We note that different off-axis angles yield different values of the 90\% enclosed-energy radii. The median aperture radius is $1.1''$, and the 25\% and 75\% percentiles are $1.0''$ and $1.4''$, respectively, albeit the radius can reach $20''$ in a handful of extreme cases.\par
We treat the remaining sources, including both undetected sources and extended ones, as upper limits, which are derived as follows.
If there are \mbox{X-ray} sources detected within $5''$, we set the upper limit to $\max\{f_\mathrm{X}+3\sigma(f_\mathrm{X}), 2f_\mathrm{X}\}$, where $f_\mathrm{X}$ and $\sigma(f_\mathrm{X})$ are the \mbox{X-ray} flux and corresponding uncertainty of the \mbox{X-ray} source(s), respectively. We set a minimum value of $2f_\mathrm{X}$ to ensure that the limit is sufficiently above $f_\mathrm{X}$ even when $\sigma(f_\mathrm{X})$ is small. This conservative choice accommodates cases where (i) the nuclear source is extended and thus was removed from our fiducial sample; (ii) one of the nearby sources is the actual \mbox{X-ray} counterpart but was missed during matching (perhaps owing to an uncertain galaxy center determination); or (iii) nearby \mbox{X-ray}-bright sources may elevate the threshold sensitivity. In most other cases, which account for 88\% of all instances, we adopt the sensitivity threshold quoted by CSC 2.1 as the actual upper limits.\par

\begin{figure}
\includegraphics[width=\hsize]{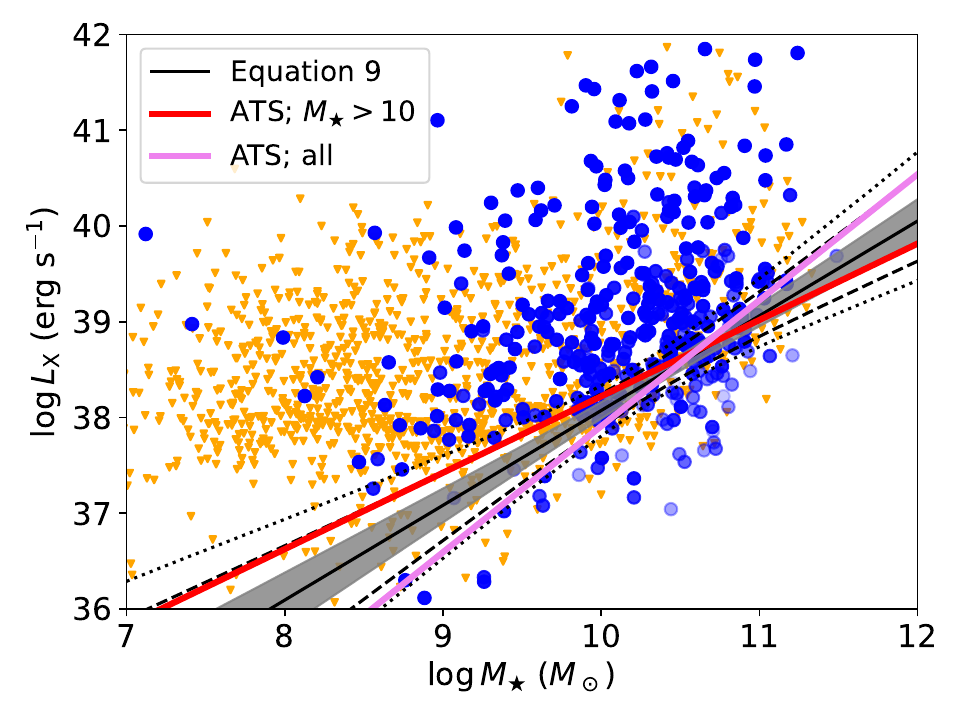}
\caption{Nuclear \mbox{X-ray} luminosity vs. $M_\star$. Blue points represent sources with compact \mbox{X-ray} nuclei, with increased color transparency reflecting an increasing likelihood of XRB contamination to the signal. Most of the blue points are not transparent because most sources have low $P_\mathrm{XRB}$. Downward orange triangles represent upper limits, which include both non detections as well as (few) extended sources. The solid red and violet lines are nonparametric, Akritas-Theil-Sen (ATS) regression lines for massive galaxies (with $M_\star>10^{10}~M_\odot$) and all the galaxies, respectively. The solid black line represents the best-fit $\log L_\mathrm{X}=\log M_\star+\log\lambda=\log M_\star+\alpha+\beta(\log M_\star-10)$ relation (Equation~\ref{eq: mu}). The shaded region, dashed lines, and dotted lines represent the $1\sigma$, $2\sigma$, and $3\sigma$ uncertainties, respectively.}
\label{fig: LxMstar}
\end{figure}

Fluxes are converted to 2-10 keV luminosities ($L_\mathrm{X}$) using the distances quoted in 50MGC, which use the best available redshift-independent and redshift-based distances. Figure~\ref{fig: LxMstar} shows the 1,606 target galaxies (out of which 388 have point-like nuclear \mbox{X-ray} sources) in the $M_\star-L_\mathrm{X}$ plane. As expected, the \mbox{X-ray} detected nuclei are much fainter than typical AGNs, whose luminosities usually exceed $10^{42}~\mathrm{erg~s^{-1}}$. Three of the \mbox{X-ray} detected nuclei belong to galaxies with $M_\star<10^8~M_\odot$ and have luminosities exceeding $L_\mathrm{X}\gtrsim10^{39}~\mathrm{erg~s^{-1}}$. All were previously reported in the literature; more details on these systems are given in Appendix~\ref{append: lowmass}.\par

The figure also presents some representative regression lines. The solid red and violet lines are based on the Akritas-Theil-Sen estimator \citep{Akritas95, Helsel11, Feigelson12}, which is a nonparametric linear regression method for censored data (i.e., including detection limits). The red line is for galaxies more massive than $M_\star=10^{10}~M_\odot$, which generally have $f_\mathrm{occ}=1$, while the violet line is for all the sources. As we will discuss further in Section~\ref{sec: method}, adopting a linear relation to depict the relation is only sensible for galaxies actually hosting MBHs--whereas a large fraction of the upper limits at $M_\star<10^{10}~M_\odot$ may be due to not hosting MBHs. Therefore, the red line is more scientifically relevant, although it relies on extrapolations down to $M_\star<10^{10}~M_\odot$; the violet line effectively represents the extreme case of $f_\mathrm{occ}=1$ across the whole $M_\star$ range. These nonparametric lines only serve as approximate presentations, whereas our parametric, scientifically motivated model will be presented in Section~\ref{sec: method}, whose corresponding relation is represented by the black line in Figure~\ref{fig: LxMstar}. The black line is roughly between the two Akritas-Theil-Sen nonparametric lines and is thus reasonable at least at face value. More detailed validations of our model assumptions will be discussed in Sections~\ref{sec: method} and \ref{sec: discussion}.

\subsection{\mbox{X-Ray} Binary Contamination Assessment}
\label{subsec: xrb}
To estimate the sample's active fraction, we must first assess, for each nuclear \mbox{X-ray} source, the likelihood that the measured \mbox{X-ray} luminosity originates from XRBs rather than from low-level accretion onto a central MBH. This issue was explored at length by \citet{Foord17} and \citet{Lee19}, so we follow their approach here. The XRB population include long-lived low-mass XRBs with donor star masses $\lesssim2~M_\odot$ and young high-mass XRBs with donor star masses $\gtrsim8~M_\odot$, and XRBs with intermediate mass donors quickly evolve into low-mass XRBs due to a strong accretion instability. For the low- and high-mass XRB populations, we adopt the X-ray luminosity functions derived by \citet{Lehmer19} and \cite{Lehmer21}, respectively. The former depends on the host galaxy's stellar mass, while the latter depends on the host star formation rate (SFR) and metallicity.  

The 50MGC provides host galaxy $M_\star$; to measure SFRs, we proceed as follows. The Local Volume Galaxy catalog\footnote{\url{https://relay.sao.ru/lv/lvgdb/}} \citep{Karachentsev13b} provides SFRs for 50MGC galaxies within $\approx11$~Mpc, measured from H$\alpha$ or far-UV fluxes \citep{Karachentsev13a}, and we adopt their results for these systems. For others, we infer SFRs by fitting their spectral energy distributions (SEDs). Photometry of nearby galaxies through typical source-detection pipelines is often less reliable due to over-segmentation into multiple components. Therefore, we rely on dedicated catalogs and adopt the far-UV to mid-infrared photometry compiled in the NASA-Sloan Atlas (NSA)\footnote{\url{https://www.sdss4.org/dr17/manga/manga-target-selection/nsa/}} and the Siena Galaxy Atlas 2020 (SGA-2020; \citealt{Moustakas23}).\footnote{\url{https://www.legacysurvey.org/sga/sga2020/}}

The 50MGC already provides matched identifiers in the NSA. For SGA-2020, most of our sources with central compact X-ray detections (304 out of 388) can be matched using a $2''$ matching radius, while unmatched sources primarily reside outside the SGA-2020 footprint. We fit the corresponding SEDs with \texttt{CIGALE} (Code Investigating GALaxy Emission) v2025.0 \citep{Boquien19, Yang20}, following the parameter settings in Table~4 of \citet{Zou22}, except that we do not include the X-ray module. Briefly, we adopt the stellar templates from \citet{Bruzual03}, the Chabrier initial mass function \citep{Chabrier03}, a delayed star-formation history, the dust attenuation model from \citet{Calzetti00}, and the dust emission templates from \citet{Dale14}. We refer readers to Section~3 of \citet{Zou22} for further discussion of the SED fitting methodology. The fitting returns both $M_\star$ and SFR. We have verified that the returned $M_\star$ values are in good agreement with the 50MGC values, with a median offset of 0.1~dex and a scatter of 0.2~dex. However, we will retain the 50MGC $M_\star$ for the remaining analyses to maintain consistency. In total, 86\% of our galaxies have SFR measurements. 
\par
Next, we turn to high-resolution photometric data (available for 60\% of the \mbox{X-ray} detected galaxies) to estimate the {\it nuclear} \mstar\ and SFR. Specifically, for those galaxies with Hubble Space Telescope (HST) data, we measure the fraction of light by fitting isophotes to HST imaging. We use observations from either the Advanced Camera for Surveys Wide Field Camera or the Wide Field Camera 3 UVIS/IR. In order to capture the underlying stellar population and mitigate contamination from star forming region, we restrict our measurements to data in wide, relatively red HST filters. Filters used included F555W, F606W, F814W, F850LP, F110W, and F160W. We calculate the enclosed light fraction, $\gamma$, within an aperture radius equal to the 90\% enclosed energy radius of the compact \mbox{X-ray} source. The nuclear $M_\star$ and SFR are then obtained by multiplying the total $M_\star$ and SFR by $\gamma$, thereby assuming a constant mass-to-light ratio.\par
We follow \citet{Lee19} to calculate the probability $P_\mathrm{XRB}$ that the expected nuclear XRB emission {\it exceeds} a given $L_\mathrm{X}$. The expected {\it number} of XRBs is $\mathbb{E}(N)=\int\frac{dN}{dL}dL$, where $\frac{dN}{dL}$ is the (\mstar- and SFR-dependent) XRB \mbox{X-ray} luminosity function. This number is a Poisson random variable, and the {\it total} XRB \mbox{X-ray} luminosity is given by:  $L_\mathrm{X}^\mathrm{XRB}=\sum_{i=1}^NL_{\mathrm{X},i}^\mathrm{XRB}$, where  $L_{\mathrm{X},i}^\mathrm{XRB}$ is the individual XRB luminosity drawn from the XRB \mbox{X-ray} luminosity function in \citet{Lehmer19, Lehmer21}. We note that this approach is notably different from, and arguably superior to, discarding as "contaminated" all nuclear \mbox{X-ray} sources with luminosity below 2 or 3 $\sigma$ times the expected luminosity arising from XRBs (e.g., \citealt{Mezcua18, Birchall20, Bykov24}) because of two reasons. First, the latter adopts a normal approximation that cannot accurately depict the tail distribution of $L_\mathrm{X}^\mathrm{XRB}$, to which $P_\mathrm{XRB}$ is sensitive. Second, much information is lost when simply discarding contaminated sources.\par
We generate $10^4$ realizations of $L_\mathrm{X}^\mathrm{XRB}$ and derive $P_\mathrm{XRB}$ with the following expression: $P_\mathrm{XRB}=P(L_\mathrm{X}^\mathrm{XRB}\ge L_\mathrm{X})$. $P_\mathrm{XRB}$ is a function of the nuclear $L_\mathrm{X}$ and the aperture radius $r$ (expressed in physical units), plus other source-specific quantities, i.e., \mstar, SFR and the surface brightness profile. We further define $\lambda=L_\mathrm{X}/M_\star$ (in units of $\mathrm{erg~s^{-1}}~M_\odot^{-1}$) to normalize $L_\mathrm{X}$ over $M_\star$ (see Section~\ref{subsec: model} for more discussions of $\lambda$). We take the $P_\mathrm{XRB}(\lambda, r; M_{\star}, {\rm SFR}, \gamma)$ functions derived above for the galaxies with HST data and calculate their median $P_\mathrm{XRB}(\lambda, r)$ separately for early and late types (the latter tends to have higher high-mass XRB contributions). We use these median functions to estimate $P_\mathrm{XRB}$ for the galaxies with no available HST data, which are assumed to follow a Sersic profile with index 4 for early types and 1 for late types. The resulting median $P_\mathrm{XRB}$ of our sample is 5\%, and the 25\% and 75\% percentiles are 1\% and 18\%, respectively. 
\par
To verify the applicability of these median $P_\mathrm{XRB}(\lambda, r)$ functions to the sources lacking HST coverage, we apply them to predict $P_\mathrm{XRB}$ for the HST-covered galaxies, which have their own $P_\mathrm{XRB}$ measurements available. We find that the predicted and the actual $P_\mathrm{XRB}$ are reasonably consistent across the $M_\star$ range -- the median values of the actual and predicted $P_\mathrm{XRB}$ for HST-covered galaxies are 2\% and 5\%, respectively. 

We note that the adopted X-ray luminosity function, and the ensuing $P_\mathrm{XRB}$, already account for the possible presence of ultraluminous X-ray sources (ULXs; see discussion in \citealt{Lehmer21}). In comparison, the scaling relation with $M_\star$ and SFR obtained by \citet{Kovlakas20} yields an average rate of 0.02 ULX with $L_\mathrm{X}>10^{39}~\mathrm{erg~s^{-1}}$ per galactic nucleus in our sample, which is comparable to our derived $P_\mathrm{XRB}$.
Although there are inevitable uncertainties when deriving $P_\mathrm{XRB}$--for example, the Sersic index of low-mass early-type galaxies may deviate from the nominal value of 4 \citep{Kormendy09}--the exact values of $P_\mathrm{XRB}$ are generally not critical. This is because they are sufficiently small, and the detected X-ray signals primarily originate from MBHs.\par
We summarize the properties of all the 1,606 sources in our sample in Table~\ref{tbl: sample}.

\begin{table*}
\caption{Source properties of the sample}
\label{tbl: sample}
\centering
\begin{threeparttable}
\begin{tabular}{cccccccc}
\hline
\hline
Name & RA & Dec & Distance & $\log L_\mathrm{X}$ & $\log M_\star$ & $\log\mathrm{SFR}$ & $P_\mathrm{XRB}$\\
& (deg) & (deg) & (Mpc) & ($\mathrm{erg~s^{-1}}$) & ($M_\odot$) & ($M_\odot~\mathrm{yr^{-1}}$)\\
(1) & (2) & (3) & (4) & (5) & (6) & (7) & (8)\\
\hline
2MASXJ15150097+5525555 & 228.753971 & 55.432026 & 49.67 & $39.22\pm0.16$ & 9.66 & $-2.79$ & 2.1\%\\
ESO005-004 & 91.423788 & $-86.631870$ & 23.66 & $40.50\pm0.04$ & 10.17 & nan & 0.6\%\\
ESO097-013 & 213.291254 & $-65.339152$ & 4.20 & $38.75\pm0.02$ & 10.06 & nan & 0.6\%\\
ESO121-006 & 91.874124 & $-61.807596$ & 14.14 & $40.37\pm0.01$ & 9.47 & $-0.16$ & 0.0\%\\
ESO138-001 & 252.834282 & $-59.234564$ & 29.37 & $41.25\pm0.01$ & 9.81 & nan & 0.0\%\\
$\cdots$ & $\cdots$ & $\cdots$ & $\cdots$ & $\cdots$ & $\cdots$ & $\cdots$ & $\cdots$\\
\hline
\hline
\end{tabular}
\begin{tablenotes}
\item
\textit{Notes.} Column (1): galaxy names as in the 50MGC. Columns (2) and (3): J2000 coordinates. Column (4): distances. Column (5): nuclear $2-10$~keV luminosities. Column (6): host-galaxy stellar masses. Column (7): host star-formation rates. Column (8): the probability that the detected nuclear \mbox{X-ray} emission is from XRBs. The table in its entirety is available online.
\end{tablenotes}
\end{threeparttable}
\end{table*}

\section{Bayesian Inference}
\label{sec: method}
It is apparent from Figures~\ref{fig: MstarDist} and \ref{fig: LxMstar} that the fraction of detected \mbox{X-ray} nuclei decreases dramatically with decreasing \mstar: 54\% of sources with $M_\star > 10^{10}~M_\odot$ have detected compact \mbox{X-ray} nuclei, while this fraction drops to 12\% for $M_\star < 10^{10}~M_\odot$. Qualitatively, this could imply that MBHs are ubiquitous in low-mass galaxies but are significantly more underluminous than the detection limits, that the MBH occupation fraction decreases at lower masses, or a combination of the two. In this section, we quantitatively constrain $f_\mathrm{occ}$. 

\subsection{Modeling Occupation Fraction and Specific Accretion-Rate Distribution}
\label{subsec: model}
Our methodology updates the framework presented in both \citetalias{Miller15} and \citetalias{Burke25}. We first define the specific accretion rate, $\lambda = L_\mathrm{X} / M_\star$, as an empirical proxy for the Eddington ratio. This choice aligns with the definition used in several recent works on AGN accretion rates (e.g., \citealt{Bongiorno16, Yang18, Zou24b, Guetzoyan25}) and aims to bypass the uncertainties introduced by using the actual Eddington ratio $\lambda_\mathrm{Edd} = k_\mathrm{bol} L_\mathrm{X} / L_{\rm Edd}$. These uncertainties include the choice of the scaling relation between $M_\star$ and $M_\mathrm{BH}$ (and its scatter) as well as the \mbox{X-ray} to bolometric correction $k_\mathrm{bol}$. For reference, $\lambda = 10^{34}~\mathrm{erg~s^{-1}}~M_\odot^{-1}$ corresponds to an Eddington ratio of $\lambda_\mathrm{Edd} \approx 1$ when assuming $k_\mathrm{bol} = 25$ and a typical scaling relation of $M_\mathrm{BH} = 0.002\,M_\star$ (as discussed in Section~\ref{subsec: conversion}, $k_\mathrm{bol}$ likely depends on both $L_\mathrm{X}$ and $\lambda_\mathrm{Edd}$; however, because $k_\mathrm{bol}$ is not included in the definition of $\lambda$, its exact value does not affect our results).
\par
We define the specific accretion-rate distribution function (sARDF) as the conditional probability density per unit $\log\lambda$ for a galaxy with a central MBH to have a specific accretion rate of $\lambda$. The sARDF is represented by $p(\lambda) = p(\log\lambda \mid M_\star, I_\mathrm{MBH} = 1)$, where $I_\mathrm{MBH} = 1$ indicates the presence of a central MBH. 
We first consider the simple case where there is no XRB contamination (i.e., $P_\mathrm{XRB}=0$) to help illustrate our methodology. As will be shown in Section~\ref{subsec: posterior}, the analysis for a non-zero $P_\mathrm{XRB}$ can be generalized from this case. Our data are described by the following distribution:
\begin{align}
\log\lambda\sim\mathrm{Mixture}\{[p(\lambda), \delta(-\infty)], [f_\mathrm{occ}, 1-f_\mathrm{occ}]\},\label{eq: base}
\end{align}
where ``Mixture'' stands for a mixture model in statistics. This means a galaxy has a probability of $f_\mathrm{occ}$ of hosting a central MBH, in which case its $\log\lambda$ follows the distribution $p(\lambda)$. Conversely, the galaxy has a probability of $(1-f_\mathrm{occ})$ of not hosting a central MBH, resulting in $\log\lambda = -\infty$. Therefore, we need to simultaneously constrain both $p(\lambda)$ and $f_\mathrm{occ}$.
\par
Since $p(\lambda)$ and $f_\mathrm{occ}$ are coupled in Equation~\ref{eq: base}, the constraints on $f_\mathrm{occ}$ depend on the choice of $p(\lambda)$.
Previous work by \citetalias{Miller15} assumes a log-linear relation with log-normal scatter between the MBH accretion luminosity in \mbox{X-rays}, $L_\mathrm{X}$, and the host-galaxy stellar mass, $M_\star$. A more recent, multiwavelength investigation by \citetalias{Burke25} assumes that the luminosity scatter at a given mass is described by the Eddington ratio distribution function (ERDF). The probability distribution of the ERDF is modeled as a mass-independent, smoothly broken power-law, implying that galaxies have an increasingly high likelihood of hosting underluminous MBHs. This choice is based on \citet{Weigel17}, who conclude that the inferred ERDF for (luminous) AGN follows a broken power-law and is roughly mass-independent in the local universe.\footnote{The mass independence is not necessarily correct in the distant universe. As shown by \citet{Aird18} and \citet{Zou24b}, the ERDF becomes increasingly mass-dependent as the redshift increases up to $z\sim2$.} The ERDF chosen by \citetalias{Burke25} introduces a key difference compared to \citetalias{Miller15}'s approach, as \citetalias{Burke25}'s ERDF is inconsistent with log-normal scatter in the $L_\mathrm{X}$:$M_\star$ relation. In this work, we will adopt a flexible distribution family for $p(\lambda)$ to encompass both the possibilities presented by \citetalias{Miller15} and \citetalias{Burke25}, and we will revisit the viability of each functional form in Section~\ref{subsec: plateu}.

\subsection{Parametrization}
\begin{figure*}
\centering
\includegraphics[width=\linewidth]{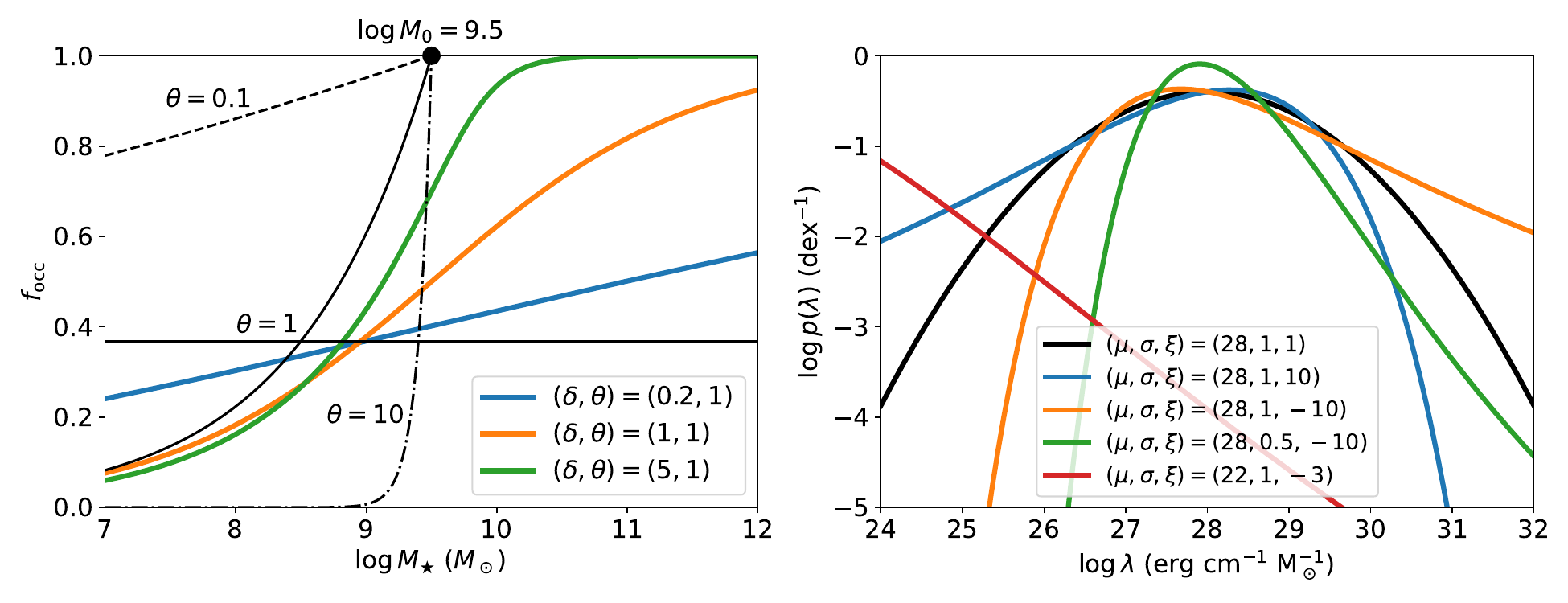}
\caption{Visual illustration of the chosen parameterization for $f_\mathrm{occ}$ (left; Equation~\ref{eq: focc}) and $p(\lambda)$ (right; Equation~\ref{eq: plambda}). \textit{Left}: In this plot, $\log M_0=9.5$. The black horizontal line represents the limiting $f_\mathrm{occ}$ as $\delta \to 0$ for any $\theta$. The other black curves correspond to $\delta \to +\infty$ at different $\theta$ values, reaching $f_\mathrm{occ}=1$ at $\log M_\star=\log M_0$. The colored curves show how $f_\mathrm{occ}$ evolves with $\delta$ when $\theta=1$. As $\delta$ increases, the curve becomes less flat and approaches the corresponding limiting black curve for $\theta=1$. \textit{Right}: The plot illustrates $p(\lambda)$ for different $(\mu, \sigma, \xi)$ values, as indicated in the legend. When $\xi=1$, $p(\lambda)$ is a normal distribution, shown by the black curve. The blue and orange curves have different $\xi$ values, resulting in heavier tails: on the left when $\xi>1$ and on the right when $\xi<1$. The green curve, compared to the orange curve, demonstrates that $\sigma$ controls the width of $p(\lambda)$. The red curve shows that our $p(\lambda)$ can mimic a power-law distribution over the plotted $\log\lambda$ range when both $\mu$ and $\xi$ are small.}
\label{fig: illustration}
\end{figure*}

To parameterize $f_\mathrm{occ}(M_\star)$ and $p(\lambda)$, we aim to adopt flexible functional forms that can empirically represent as broad a range of behaviors as possible. For $f_\mathrm{occ}$, we follow \citetalias{Burke25} and assume that the expression parameterizing the MBH occupation fraction as a function of $M_\star$ is a generalized logistic function. This family of functions is sufficiently flexible to model typical growth patterns \citep{Richards59}. It has the form:
\begin{align}
f_\mathrm{occ}(M_\star)=\left[1+\delta e^{-\theta\delta\left(\log M_\star-\log M_0\right)}\right]^{-\frac{1}{\delta}},\label{eq: focc}
\end{align}
where $\log M_0$, $\delta$, and $\theta$ are free parameters. A minor difference between our parameterization and that of \citetalias{Burke25} is the inclusion of an additional $\delta$ factor in the exponentiation term, which reduces the correlation between $\theta$ and $\delta$. In our definition, $\theta$ controls the limiting behavior of $f_\mathrm{occ}$ with respect to $\delta$, effectively decoupling from $\delta$ at large values. Mathematically, it can be shown that:
\begin{align}
\lim_{\delta\to0}f_\mathrm{occ}&=\frac{1}{e}\\
\lim_{\delta\to+\infty}f_\mathrm{occ}&=\min\{e^{\theta(\log M_\star-\log M_0)}, 1\}
\end{align}
The above equation indicates that $\theta$ (when positive) directly affects $f_\mathrm{occ}$ in low-mass galaxies, particularly when $\delta \gg 1$. The left panel of Figure~\ref{fig: illustration} illustrates how these parameters influence the shape of $f_\mathrm{occ}$. In brief, $\log M_0$ shifts the function along the $\log M_\star$ axis. Increasing $\theta$ suppresses $f_\mathrm{occ}$ when $\delta \gg 1$, and increasing $\delta$ makes the curve less flat and closer to the limiting $f_\mathrm{occ}(\delta \to +\infty)$ curve.

For the parameterization of $p(\lambda)$, we assume that it follows a Box-Cox distribution for a given $M_\star$, and that the distribution can shift linearly with $\log M_\star$. This family of distributions includes commonly adopted models for the Eddington ratio or specific accretion rate distribution, such as the normal distribution (\citetalias{Miller15}) and the power-law distribution (\citetalias{Burke25}). The Box-Cox distribution, also known as the power-normal distribution (e.g., \citealt{Freeman06}), originates from the Box-Cox transformation \citep{Box64}:
\begin{align}
Y=\frac{X^\xi-1}{\xi},~~X>0, \label{eq: boxcoxtrans}
\end{align}
which reduces to $Y=\ln X$ for $\xi=0$. The Box-Cox transformation is a statistical technique used to stabilize variance and make data more closely approximate a normal distribution. It is particularly useful for transforming skewed data $X$, often with sizable tails, into a more symmetrical bell-shaped distribution for $Y$. In our model, we will use $X=\log\lambda$ and set $\xi$ as a free parameter. If $Y$ follows a (truncated\footnote{$Y$ is truncated by definition because $X$ is assumed to be positive. Specifically, $Y>-\frac{1}{\xi}$ for $\xi>0$, and $Y<-\frac{1}{\xi}$ for $\xi<0$. Truncation is almost negligible in our modeling because $X=\log\lambda \gg 0$.}) normal distribution, then $X=\log\lambda$ follows a Box-Cox distribution. Note that $Y = X - 1$ if $\xi = 1$, making a truncated normal distribution a special case of the Box-Cox distribution. Additionally, for certain values of $\xi$, a Box-Cox distribution can model a power-law-like tail over a large range of $X$.\par
We assume the following truncated normal distribution for $Y$: 
\begin{align}
Y\sim
\begin{cases}
\mathcal{TN}(\frac{\mu^\xi-1}{\xi}, \mu^{2\xi-2}\sigma^2; -\frac{1}{\xi}, +\infty), \xi>0\\
\mathcal{TN}(\frac{\mu^\xi-1}{\xi}, \mu^{2\xi-2}\sigma^2; -\infty, -\frac{1}{\xi}), \xi<0
\end{cases},\label{eq: Y}
\end{align}
where $\mu$ and $\sigma$ are parameters that approximate, but are not identical to, the mean and standard deviation of $\log\lambda$, and $Y\sim\mathcal{TN}(a_1, a_2; a_3, a_4)$ represents a normal distribution with a mean of $a_1$ and a variance of $a_2$, constrained within the interval $a_3<Y<a_4$. \par
Inverting the transformation in Equation~\ref{eq: boxcoxtrans}, we obtain the following expression for the sARDF: 
\begin{align}
p(\lambda)&=\frac{1}{\sqrt{2\pi}\sigma A}\left(\frac{\log\lambda}{\mu}\right)^{\xi-1}e^{-\frac{1}{2}\left(\frac{\mu}{\xi\sigma}\right)^2\left[\left(\frac{\log\lambda}{\mu}\right)^\xi-1\right]^2},\label{eq: plambda}\\
A&=\Phi\left(\frac{\mu}{|\xi|\sigma}\right),\label{eq: A}
\end{align}
where $\mathrm{sgn}(x)$ is the sign function, and $\Phi(x)=\frac{1}{2}\left[1+\mathrm{erf}\left(\frac{x}{\sqrt{2}}\right)\right]$ is the cumulative distribution function of the standard normal distribution. 

We illustrate how these parameters influence $p(\lambda)$ in the right panel of Figure~\ref{fig: illustration}. Changing $\mu$ shifts the curve, while $\sigma$ controls the width of $p(\lambda)$. For a given set of $(\mu, \sigma)$, $\xi$ determines the overall shape of $p(\lambda)$. When $\xi=1$, $p(\lambda)$ is a normal distribution, as shown by the black curve in the figure. As $\xi$ increases (or decreases), $p(\lambda)$ develops a heavier left (or right) tail, as illustrated by the blue and orange curves. A notable case occurs when $p(\lambda)$ mimics a power-law distribution, represented by the red curve; in this case, both $\mu$ and $\xi$ are small, and the distribution approximates a power-law across the plotted range of $\log\lambda$.
To account for potential changes in $p(\lambda)$ with $M_\star$, we further assume that $\mu$ is linearly correlated with $\log M_\star$:
\begin{align}
\mu=\alpha+\beta(\log M_\star-10),\label{eq: mu}
\end{align}
where $\alpha$ and $\beta$ are free coefficients. This allows $p(\lambda)$ to shift linearly with $\log M_\star$ while maintaining roughly the same overall shape for different $M_\star$ values.

\subsection{Likelihood}
\label{subsec: posterior}
We explicitly derive the likelihood from Equation~\ref{eq: base} below. We first present the likelihood for the case when the data are noiseless, and then we derive the full likelihood with a rigorous treatment of the measurement uncertainties. The former is relatively simpler and helps to illustrate the model, while the latter, although much more complex, can intuitively be regarded as a perturbative correction to the noiseless likelihood. Hereafter, we use the subscript or superscript $k$ to refer to a given galaxy with the identifier $k$ in our sample.

\subsubsection{Noiseless Likelihood}
\label{sebsec: like_noiseless}
For an uncensored (i.e., compact \mbox{X-ray} detected) source $k$, the likelihood is given by: 
\begin{align}
\ln\mathcal{L}_k^\mathrm{det}=\ln p(\lambda_k)+\ln f_\mathrm{occ}(M_k).\label{eq: Ldet_noiseless}
\end{align}
For censored sources (i.e., with $L_\mathrm{X}$ upper limits), Equation~\ref{eq: base} reduces to
\begin{align}
(I_k=0)&\sim\mathrm{Bernoulli}(p_k),\\
p_k&=f_\mathrm{occ}(M_k)S_\lambda(\log\lambda_{\mathrm{lim},k}),
\end{align}
where $I_k$ is set to 0[/1] if $\lambda_k<\lambda_{\mathrm{lim},k}$ [/$\lambda_k\ge\lambda_{\mathrm{lim},k}$]; $\lambda_\mathrm{lim}$ is the upper limit of $\lambda$; $p_k$ is the probability that a source $k$ has $\lambda_k\ge\lambda_{\mathrm{lim},k}$; and $S_\lambda(\log\lambda_\mathrm{lim})=P(\{\log\lambda\sim p(\lambda)\}\ge\log\lambda_\mathrm{lim})$ is the survival function corresponding to $p(\lambda)$. 

$S_\lambda(\log\lambda_\mathrm{lim})$ can be derived from Equations~\ref{eq: boxcoxtrans} and \ref{eq: Y}:
\begin{align}
S_\lambda(\log\lambda_\mathrm{lim})=
\begin{cases}
\frac{1}{A}\left[1-\Phi\left(\frac{\mu}{\xi\sigma}\left[\left(\frac{\log\lambda_\mathrm{lim}}{\mu}\right)^\xi-1\right]\right)\right], \xi>0\\
1-\frac{1}{A}\Phi\left(\frac{\mu}{\xi\sigma}\left[\left(\frac{\log\lambda_\mathrm{lim}}{\mu}\right)^\xi-1\right]\right), \xi<0
\end{cases}.\label{eq: Slambda}
\end{align}
The likelihood for a censored source $k$ is given by: 
\begin{align}
\ln\mathcal{L}_k^\mathrm{cens}=\ln(1-p_k).
\end{align}\par
Note that the above equations hold when $P_\mathrm{XRB}=0$, as assumed in Equation~\ref{eq: base}. If $P_\mathrm{XRB}\neq 0$, the log-likelihood function of an uncensored source is the weighted mean of $\ln\mathcal{L}^\mathrm{det}$ and $\ln\mathcal{L}^\mathrm{cens}$:
\begin{align}
\ln\mathcal{L}_k=(1-P_{\mathrm{XRB},k})\ln\mathcal{L}_k^\mathrm{det}+P_{\mathrm{XRB},k}\ln\mathcal{L}_k^\mathrm{cens},\label{eq: nonzero}
\end{align}
where the first term refers to sources whose \mbox{X-ray} emission originates solely from MBH accretion, and the second term addresses sources whose \mbox{X-ray} emission is contaminated by XRBs. Consequently, the overall log-likelihood function is: 
\begin{align}
\ln\mathcal{L}=&\sum_{k\in\{\mathrm{uncensored}\}}\left[(1-P_{\mathrm{XRB},k})\ln\mathcal{L}_k^\mathrm{det}+P_{\mathrm{XRB},k}\ln\mathcal{L}_k^\mathrm{cens}\right]\nonumber\\
&+\sum_{k\in\{\mathrm{censored}\}}\ln\mathcal{L}_k^\mathrm{cens}.\label{eq: lnL_full}
\end{align}

\subsubsection{Likelihood with Measurement Uncertainties}
\label{subsec: like_final}
We now extend the derivation by incorporating measurement uncertainties into the likelihood function. This allows for a more realistic and rigorous statistical treatment of the data.
Our data include uncertainties in both axes ($\log M_\star$ and $\log\lambda$), and it is generally a nontrivial task to account for uncertainties in both variables, especially for the abscissa. We derive the corresponding likelihood following the Bayesian approach described in \citet{Kelly07}. The key idea is to treat the true values of $(\log M_\star, \log\lambda)$--which would be available in the case of perfectly precise measurements--as missing data, and then introduce and marginalize over these latent parameters for each data point. We denote $(\chi, \zeta)$ as the unknown true values of $(\log M_\star, \log\lambda)$. The likelihood for each source $k$ is:
\begin{align}
\mathcal{L}_k=\iint p(\log M_k, \log\lambda_k\mid\chi_k,\zeta_k)p(\zeta_k\mid\chi_k)p(\chi_k)d\chi_kd\zeta_k.\label{eq: Lk_precise}
\end{align}
Throughout this subsection, we denote $\mathcal{N}(\vec{m}, \Sigma)$ as either the univariate or multidimensional normal distribution with a mean vector $\vec{m}$ and covariance matrix $\Sigma$. We also denote $\mathcal{N}(\vec{x}\mid\vec{m}, \Sigma)$ as the corresponding probability distribution function evaluated at $\vec{x}$.\par
Following the same approach as in Section~\ref{sebsec: like_noiseless}, we first consider uncensored sources under the $p_\mathrm{XRB}=0$ condition, i.e., corresponding to $\mathcal{L}_k^\mathrm{det}$ in Equation~\ref{eq: Ldet_noiseless}. The first term of the integrand in Equation~\ref{eq: Lk_precise}, $p(\log M_k, \log\lambda_k\mid\chi_k,\zeta_k)$, represents measurement uncertainties, such that:
\begin{align}
&(\log M_k, \log\lambda_k)\mid(\chi_k, \zeta_k)\sim\mathcal{N}((\chi_k, \zeta_k), \Sigma_k),\label{eq: dist_M_lambda}\\
\Sigma_k&=\begin{pmatrix}
\Sigma_{11,k} & \Sigma_{12,k}\\
\Sigma_{12,k} & \Sigma_{22,k}
\end{pmatrix}\nonumber\\
&=\begin{pmatrix}
\sigma^2\{\log M_k\} & -\sigma^2\{\log M_k\mid D_k\}\\
-\sigma^2\{\log M_k\mid D_k\} & \sigma^2\{\log f_{X,k}\}+\sigma^2\{\log M_k\mid D_k\}
\end{pmatrix}
\end{align}
where $\sigma\{\log M_k\}$ and $\sigma\{\log f_{X,k}\}$ are the uncertainties in $\log M_k$ and $\log f_\mathrm{X}$, respectively, and $D$ is the source distance. The $\sigma\{\log M_k\}$ values quoted by \citet{Ohlson24} already include the (generally subdominant) distance uncertainties; the uncertainty contributed by the remaining, distant-independent factors is estimated to be $\sigma\{\log M_k\mid D_k\}=0.136$ (see Section~4 in \citealt{Ohlson24}). $\log\lambda$ is not affected by distance uncertainties because the distance dependence of both $L_\mathrm{X}$ and $M_\star$ canceled out. Therefore, only $\sigma\{\log M_k\mid D_k\}$ contributes to $\Sigma_{12,k}=\mathrm{Cov}\{\log\lambda_k,\log M_k\mid\chi_k, \zeta_k\}$ and $\Sigma_{22,k}=\mathrm{Var}\{\log\lambda_k\mid\chi_k, \zeta_k\}$.

The second term of the integrand in Equation~\ref{eq: Lk_precise}, $p(\zeta_k\mid\chi_k)$, is set by Equation~\ref{eq: base}:
\begin{align}
\zeta_k\mid\chi_k\sim\text{Mixture}\{[p_\lambda(\zeta_k\mid\chi_k), \delta(-\infty)], [f_\mathrm{occ}(\chi_k), 1-f_\mathrm{occ}(\chi_k)]\},
\end{align}
where $p_\lambda(\zeta_k\mid\chi_k)$ is the same as in Equation~\ref{eq: plambda} but with $(\chi_k, \zeta_k)$ replacing $(\log M_\star, \log\lambda)$. Therefore, similarly to Equation~\ref{eq: Ldet_noiseless}, we have:
\begin{align}
p(\zeta_k\mid\chi_k)=p_\lambda(\zeta_k\mid\chi_k)f_\mathrm{occ}(\chi_k).
\end{align}\par
The third term of the integrand in Equation~\ref{eq: Lk_precise}, $p(\chi_k)$, represents the underlying distribution of $\chi_k$. Following \citet{Kelly07}, we model it with a mixture of normal distributions:
\begin{align}
p(\chi_k)&=\sum_jw_j\mathcal{N}(\chi_k\mid t_j, s_j^2),\\
\sum_jw_j&=1.
\end{align}
In principle, the $(w_j, t_j, s_j)$ parameters could be incorporated as free parameters within the occupation fraction model. However, this would introduce a substantial number of additional parameters (e.g., eight more for a three-component mixture), significantly increasing the model’s dimensionality and computational cost. Since our sample size is very large, the $\log M_\star$ distribution is already well constrained, so fixing $(w_j, t_j, s_j)$ has negligible impact on the inference and greatly improves computational efficiency. Since Equation~\ref{eq: dist_M_lambda} gives $\log M_k\mid\chi_k\sim\mathcal{N}(\chi_k, \Sigma_{11,k})$, we have that:
\begin{align}
p(\log M_k)&=\int p(\log M_k\mid\chi_k)p(\chi_k)d\chi_k\nonumber\\
&=\sum_j\frac{w_j}{\sqrt{2\pi(\Sigma_{11,k}+s_j^2)}}e^{-\frac{(t_j-\log M_k)^2}{2(\Sigma_{11,k}+s_j^2)}}.\label{eq: plogMk}
\end{align}
By maximizing $\prod_k p(\log M_k)$ to fit the inferred $\log M_\star$ distribution for our sample (including both censored and uncensored data), we can constrain $(w_j, t_j, s_j)$. We find that a mixture of three normal distributions models the $\log M_\star$ distribution well, with the corresponding parameters $(w_j, t_j, s_j)$ given by $(0.70, 8.75, 0.97)$, $(0.17, 10.06, 0.35)$, and $(0.13, 10.64, 0.25)$.\par
With these three terms, Equation~\ref{eq: Lk_precise} becomes
\begin{align}
\mathcal{L}_k^\mathrm{det}=\sum_jw_j\iint&\mathcal{N}((\log M_k, \log\lambda_k)\mid(\chi_k, \zeta_k), \Sigma_k)\times\nonumber\\
&\mathcal{N}(\chi_k\mid t_j, s_j^2)p_\lambda(\zeta_k\mid\chi_k)f_\mathrm{occ}(\chi_k)d\chi_kd\zeta_k.
\end{align}
After solving for the first two terms of the integrand above, $\mathcal{L}_k^\mathrm{det}$ reduces to
\begin{align}
\mathcal{L}_k^\mathrm{det}&=\sum_j\frac{w_j}{\sqrt{2\pi(\Sigma_{11,k}+s_j^2)}}e^{-\frac{(t_j-\log M_k)^2}{2(\Sigma_{11,k}+s_j^2)}}\times\nonumber\\
&\iint\mathcal{N}((x_1, x_2)\mid(c_1^{kj}, c_2^{kj}), \eta^{kj})g^k(x_1, x_2)dx_1dx_2,\label{eq: Ldet_reduced_precise}\\
c_1^{kj}&=\frac{\Sigma_{11,k}}{\Sigma_{11,k}+s_j^2}(t_j-\log M_k),\\
c_2^{kj}&=\frac{\Sigma_{12,k}}{\Sigma_{11,k}+s_j^2}(t_j-\log M_k),\\
\eta^{kj}&=
\begin{pmatrix}
\eta_{11}^{kj} & \eta_{12}^{kj}\\
\eta_{12}^{kj} & \eta_{22}^{kj}
\end{pmatrix}=\begin{pmatrix}
\frac{\Sigma_{11,k}s_j^2}{\Sigma_{11,k}+s_j^2} & \frac{\Sigma_{12,k}s_j^2}{\Sigma_{11,k}+s_j^2}\\
\frac{\Sigma_{12,k}s_j^2}{\Sigma_{11,k}+s_j^2} & \frac{\Sigma_{11,k}\Sigma_{22,k}-\Sigma_{12,k}^2+\Sigma_{22,k}s_j^2}{\Sigma_{11,k}+s_j^2}
\end{pmatrix}\\
g^k(x_1, x_2)&=p_\lambda(\log\lambda_k+x_2\mid\log M_k+x_1)f_\mathrm{occ}(\log M_k+x_1).
\end{align}
There is an important practical difference between our fitting procedure and typical regressions that assume Gaussian scatter. In the latter case, $g^k(x_1, x_2)$ is a Gaussian function, so the integration in Equation~\ref{eq: Ldet_reduced_precise} can be solved analytically in closed form; previous treatments of measurement uncertainties usually rely implicitly on this property (e.g., \citealt{Kelly07, Hogg10}). However, the integration in our $\mathcal{L}_k^\mathrm{det}$ cannot be solved analytically. Accurate calculation would require a two-dimensional numerical integration for every data point during each evaluation of $\mathcal{L}_k^\mathrm{det}$, which is not practically feasible for sampling, as numerical integrations are extremely slow. We therefore seek an approximate method to avoid numerical integration.
\par
We note that the typical measurement uncertainty ($\text{median}\{\sqrt{\Sigma_{11,k}}\}=0.14$ and $\text{median}\{\sqrt{\Sigma_{22,k}}\}=0.19$) is small compared to the overall span of our data range (several dex). Therefore, in Equation~\ref{eq: Lk_precise}, the main contribution to the integrand comes from the region around $(\log M_k, \log\lambda_k)$, due to the first term, $p(\log M_k, \log\lambda_k\mid\chi_k, \zeta_k)$. The remaining part of the integrand, $p(\zeta_k\mid\chi_k)p(\chi_k)$, varies much more slowly on a scale comparable to the total data range. Thus, it is a good approximation to perform a Taylor expansion of $p(\zeta_k\mid\chi_k)$ around $(\log M_k, \log\lambda_k)$, i.e., to expand $g^k(x_1, x_2)$ around $(0, 0)$:
\begin{align}
g^k(x_1, x_2)\approx\sum_{i_1}\sum_{i_2}g_{i_1i_2}^kx_1^{i_1}x_2^{i_2},
\end{align}
where the coefficients $g_{i_1i_2}^k$ are calculated with \texttt{TaylorSeries.jl} \citep{Benet19}. The integration in Equation~\ref{eq: Ldet_reduced_precise} reduces to evaluating moments of the normal distribution $\mathcal{N}((c_1^{kj}, c_2^{kj}), \eta^{kj})$, which are given by Isserlis's theorem \citep{Isserlis18}. We expand $g^k(x_1, x_2)$ up to the fourth order, and Equation~\ref{eq: Ldet_reduced_precise} then becomes:
\begin{align}
\mathcal{L}_k^\mathrm{det}\approx&\sum_j\frac{w_j}{\sqrt{2\pi(\Sigma_{11,k}+s_j^2)}}e^{-\frac{(t_j-\log M_k)^2}{2(\Sigma_{11,k}+s_j^2)}}\times\nonumber\\
&\{g_{00}^k+g_{10}^kc_1^{kj}+g_{01}^kc_2^{kj}+g_{20}^k[(c_1^{kj})^2+\eta_{11}^{kj}]\nonumber\\
&+g_{11}^k(c_1^{kj}c_2^{kj}+\eta_{12}^{kj})+g_{02}^k[(c_2^{kj})^2+\eta_{22}^{kj}]\nonumber\\
&+g_{30}^k[(c_1^{kj})^3+3c_1^{kj}\eta_{11}^{kj}]\nonumber\\
&+g_{21}^k[(c_1^{kj})^2c_2^{kj}+2c_1^{kj}\eta_{12}^{kj}+c_2^{kj}\eta_{11}^{kj}]\nonumber\\
&+g_{12}^k[c_1^{kj}(c_2^{kj})^2+2c_2^{kj}\eta_{12}^{kj}+c_1^{kj}\eta_{22}^{kj}]\nonumber\\
&+g_{03}^k[(c_2^{kj})^3+3c_2^{kj}\eta_{22}^{kj}]\nonumber\\
&+g_{40}^k[(c_1^{kj})^4+6(c_1^{kj})^2\eta_{11}^{kj}+3(\eta_{11}^{kj})^2]\nonumber\\
&+g_{31}^k[(c_1^{kj})^3c_2^{kj}+3(c_1^{kj})^2\eta_{12}^{kj}+3c_1^{kj}c_2^{kj}\eta_{11}^{kj}+3\eta_{11}^{kj}\eta_{12}^{kj}]\nonumber\\
&+g_{22}^k[(c_1^{kj})^2(c_2^{kj})^2+(c_1^{kj})^2\eta_{22}^{kj}+(c_2^{kj})^2\eta_{11}^{kj}\nonumber\\
&~~~~~~~~~~+4c_1^{kj}c_2^{kj}\eta_{12}^{kj}+\eta_{11}^{kj}\eta_{22}^{kj}+2(\eta_{12}^{kj})^2]\nonumber\\
&+g_{13}^k[c_1^{kj}(c_2^{kj})^3+3(c_2^{kj})^2\eta_{12}^{kj}+3c_1^{kj}c_2^{kj}\eta_{22}^{kj}+3\eta_{12}^{kj}\eta_{22}^{kj}]\nonumber\\
&+g_{04}^k[(c_2^{kj})^4+6(c_2^{kj})^2\eta_{22}^{kj}+3(\eta_{22}^{kj})^2]\}.\label{eq: Lkdet_approx}
\end{align}\par
For censored sources, the likelihood is
\begin{align}
\mathcal{L}_k^\mathrm{cens}&=p(I_k=0, \log M_k)=p(\log M_k)-p(I_k=1, \log M_k).\label{eq: lnLkcens_step1_precise}
\end{align}
$p(\log M_k)$ is given in Equation~\ref{eq: plogMk}, and the second term is
\begin{align}
&p(I_k=1, \log M_k)\nonumber\\
&=\iint p(I_k=1\mid\zeta_k)p(\log M_k\mid\chi_k)p(\zeta_k\mid\chi_k)p(\chi_k)d\chi_kd\zeta_k\nonumber\\
&=\int p(\log M_k\mid\chi_k)p(\chi_k)S_\lambda(\log\lambda_{\mathrm{lim},k}\mid\chi_k)f_\mathrm{occ}(\chi_k)d\chi_k\nonumber\\
&=\sum_jw_j\int\mathcal{N}(0; x, \Sigma_{11})\mathcal{N}(x; t_j-\log M_k, s_j^2)h(x)dx\nonumber\\
&=\sum_j\frac{w_j}{\sqrt{2\pi(\Sigma_{11,k}+s_j^2)}}e^{-\frac{(t_j-\log M_k)^2}{2(\Sigma_{11,k}+s_j^2)}}\int\mathcal{N}(x; c_1^{kj}, \eta_{11}^{kj})h^k(x)dx,\label{eq: lnLkcens_step2_precise}\\
&h^k(x)=S_\lambda(\log\lambda_{\mathrm{lim},k}\mid\log M_k+x)f_\mathrm{occ}(\log M_k+x).
\end{align}
We expand $h^k(x)$ up to the fourth order:
\begin{align}
h^k(x)\approx\sum_{i=1}^4h_i^kx^i.
\end{align}
Equation~\ref{eq: lnLkcens_step1_precise} then reduces to
\begin{align}
\mathcal{L}_k^\mathrm{cens}&\approx\sum_j\frac{w_j}{\sqrt{2\pi(\Sigma_{11,k}+s_j^2)}}e^{-\frac{(t_j-\log M_k)^2}{2(\Sigma_{11,k}+s_j^2)}}\times\nonumber\\
&\{1-h_0^k-h_1^kc_1^{kj}-h_2^k[(c_1^{kj})^2+\eta_{11}^{kj}]-h_3^k[(c_1^{kj})^3+3c_1^{kj}\eta_{11}^{kj}]\nonumber\\
&-h_4^k[(c_1^{kj})^4+6(c_1^{kj})^2\eta_{11}^{kj}+3(\eta_{11}^{kj})^2]\}\label{eq: Lkcens_approx}
\end{align}\par
\begin{figure}
\includegraphics[width=\hsize]{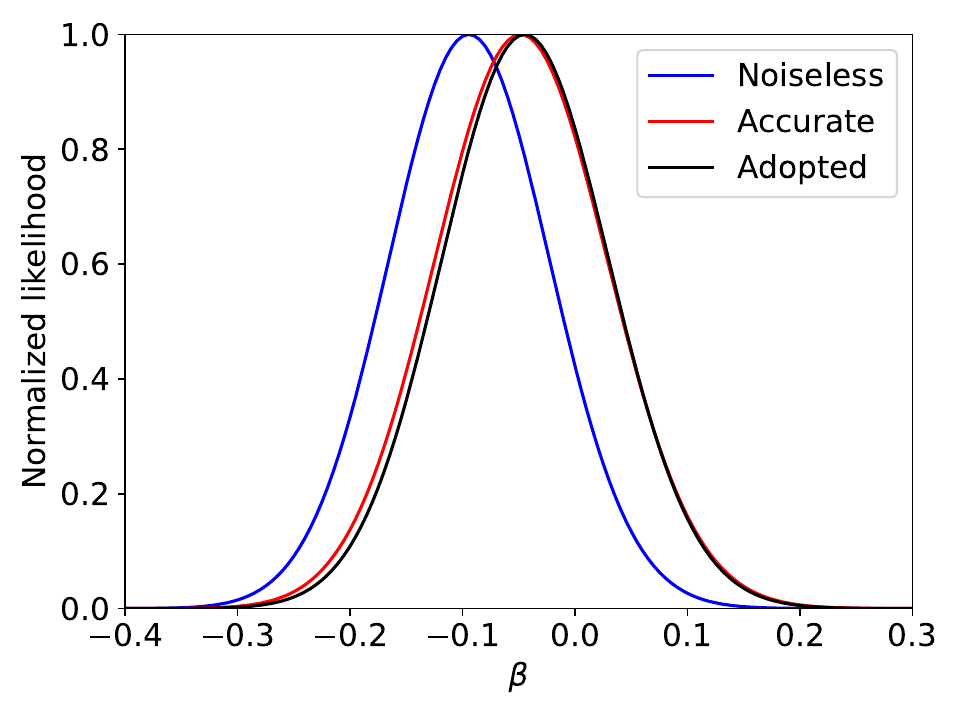}
\caption{Example comparison among the likelihood ignoring measurement uncertainties (blue), the likelihood accurately accounting for measurement uncertainties (red), and our adopted approximation (black). All likelihoods are normalized to a maximum value of one. Here, $\beta$ is varied while all other parameters are fixed at their median posterior values. The near-complete overlap of the black and red curves indicates that our adopted likelihood closely approximates the fully accurate likelihood.}
\label{fig: comp_like}
\end{figure}

The final likelihood is calculated by substituting Equations~\ref{eq: Lkdet_approx} and \ref{eq: Lkcens_approx} into Equation~\ref{eq: lnL_full}. As an example, Figure~\ref{fig: comp_like} compares the noiseless likelihood described in Section~\ref{sebsec: like_noiseless}, the precise likelihood calculated via numerical integration of Equations~\ref{eq: Ldet_reduced_precise} and \ref{eq: lnLkcens_step2_precise}, and our adopted approximate likelihood. We show how each likelihood evolves with respect to $\beta$, while holding other parameters fixed at their median posterior values from Section~\ref{subsec: sampling}. The figure demonstrates that measurement uncertainties induce a slight shift in the likelihood, and that our approximation closely matches the true likelihood, accurately accounting for measurement uncertainties. By comparing these likelihoods, we further find that $\beta$ is the parameter mostly affected by measurement uncertainties, while the $f_\mathrm{occ}$ parameters, $(\log M_0, \delta, \theta)$, generally remain much more robust against these uncertainties.

\begin{figure*}
\includegraphics[width=\hsize]{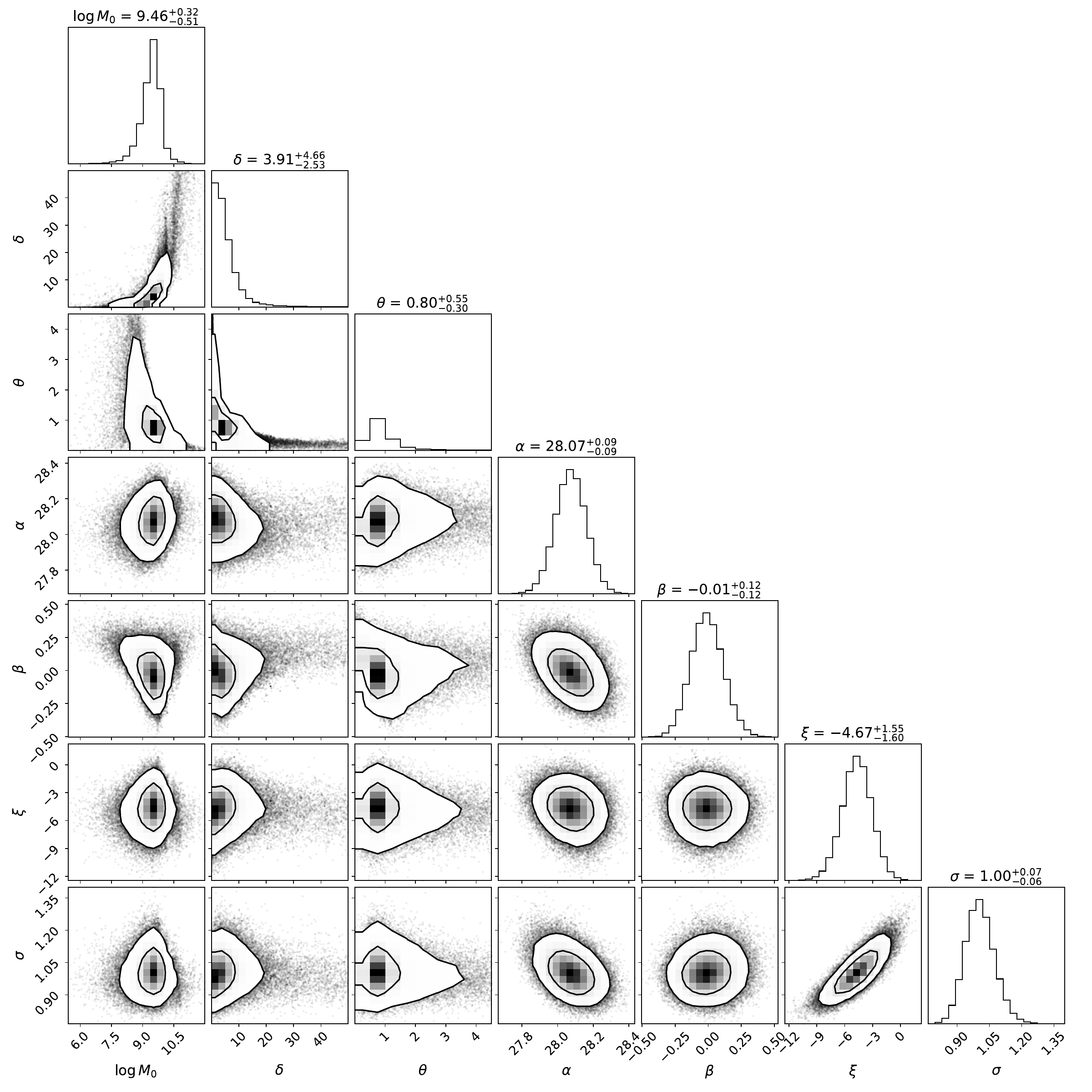}
\caption{Sampling results. Contours represent 68\% and 95\% levels; grayscale pixels represent probabilities, and the rasterized points outside the 95\% contours are individual sampling points. The first three parameters, $(\log M_0, \delta, \theta)$, represent the dependence of $f_\mathrm{occ}$ on $M_\star$ (Equation~\ref{eq: focc}), as illustrated in the left panel of Figure~\ref{fig: illustration}. $\alpha$ represents the typical $\log\lambda$ of galaxies with $M_\star=10^{10}~M_\odot$, and $\beta$ is the slope of the mass dependence of the sARDF (Equation~\ref{eq: mu}). $\xi$ and $\sigma$ control the shape of $p(\lambda)$ (Equation~\ref{eq: plambda}), as shown in the right panel of Figure~\ref{fig: illustration}. The fitted $\beta$ is consistent with zero, indicating that $p(\lambda)$ has little dependence on $M_\star$. The fitted $\xi$ is $\ll1$, suggesting that $p(\lambda)$ is highly right-skewed compared to normal distributions.}
\label{fig: sampling}
\end{figure*}
\subsection{Sampling Results}
\label{subsec: sampling}
\begin{figure*}
\centering{\includegraphics[width=0.67\hsize]{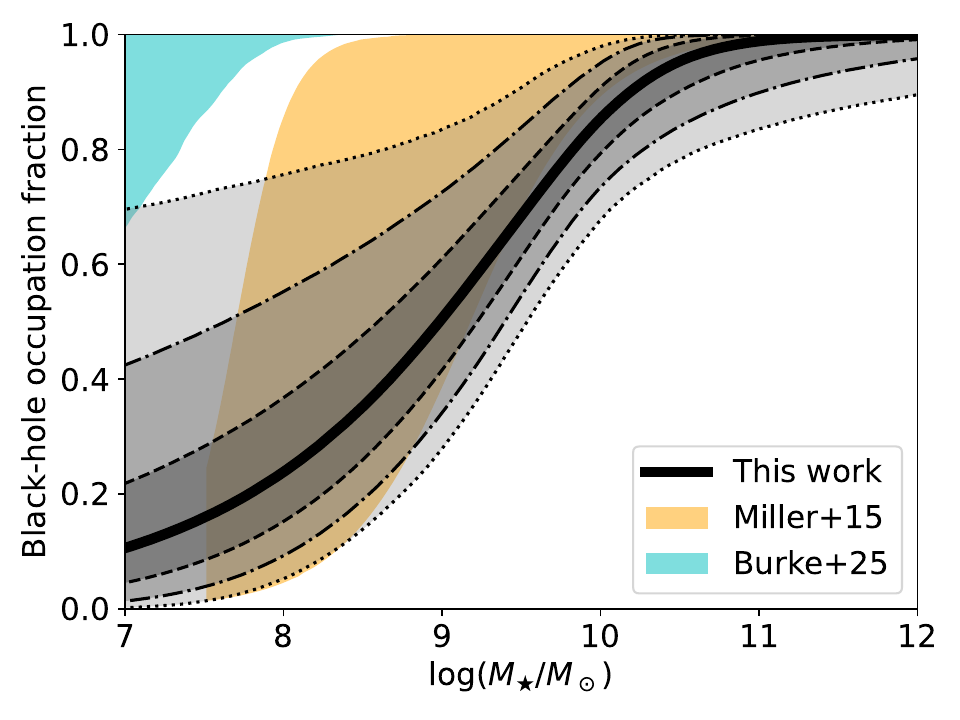}}
\caption{Local black-hole occupation fraction constraints from Chandra ACIS imaging observations of 1,606 galaxies within 50 Mpc. The black solid curve is the median of our sampling results; the dashed, dash-dotted, dotted curves and their corresponding shaded regions enclose the 1, 2, and $3\sigma$ ranges, respectively. The $1\sigma$ measurements in \citetalias{Miller15} and \citetalias{Burke25} are also shown for comparison. Our $f_\mathrm{occ}$ quickly drops down when $M_\star$ decreases below $10^{10}~M_\odot$ and is below unity at a statistically significant level for low-mass galaxies.
}
\label{fig: focc}
\end{figure*}
We assume a flat prior for all our parameters, with bounds established to ensure the propriety of the prior \citep{Tak18}: $\log M_0\sim\mathcal{U}[5, 12]$, $\delta\sim\mathcal{U}[0, 50]$, $\theta\sim\mathcal{U}[0, 10]$, $\alpha\sim\mathcal{U}[15, 40]$, $\beta\sim\mathcal{U}[-3, 3]$, $\xi\sim\mathcal{U}[-15, 15]$, and $\sigma\sim\mathcal{U}[0, 10]$. We utilize \texttt{DynamicHMC.jl},\footnote{\url{https://www.tamaspapp.eu/DynamicHMC.jl/stable/}} which implements the Hamiltonian Monte Carlo algorithm \citep{Betancourt17}, to sample the posterior distribution. The sampling results are displayed in Figure~\ref{fig: sampling}, where the median values and $1\sigma$ uncertainties for all parameters are listed at the top of each column. The fitted posterior distributions are much narrower than the prior distributions for all these parameters, indicating that the data are sufficiently informative and the inference is insensitive to the choice of prior.\par
For replicability purposes, the resulting $p(\lambda)$ can be obtained by substituting these median values of our sampled parameters into Equation~\ref{eq: plambda}.\footnote{Strictly speaking, the median $p(\lambda)$ is not exactly the same as this calculation, but we have verified that these two are nearly identical.} The results for the MBH occupation fraction are presented in Table~\ref{tbl: focc} and Figure~\ref{fig: focc}, where the median, along with the 1, 2, and 3$\sigma$ ranges of the posterior distribution, are plotted as a function of \mstar.\par
\begin{table*}
\caption{Black-hole occupation fraction as a function of host-galaxy stellar mass.}
\label{tbl: focc}
\centering
\begin{threeparttable}
\begin{tabular}{cccccccc}
\hline
\hline
$\log M_\star$ & $0.13\%$ & $2.3\%$ & $16\%$ & $50\%$ & $84\%$ & $97.7\%$ & $99.87\%$\\
(1) & (2) & (3) & (4) & (5) & (6) & (7) & (8)\\
\hline
7.0 & 0.002 & 0.014 & 0.046 & 0.106 & 0.218 & 0.425 & 0.695\\
7.1 & 0.003 & 0.017 & 0.052 & 0.115 & 0.230 & 0.436 & 0.701\\
7.2 & 0.005 & 0.021 & 0.059 & 0.125 & 0.242 & 0.447 & 0.705\\
7.3 & 0.007 & 0.026 & 0.066 & 0.136 & 0.255 & 0.459 & 0.712\\
7.4 & 0.010 & 0.032 & 0.075 & 0.147 & 0.269 & 0.471 & 0.718\\
$\cdots$ & $\cdots$ & $\cdots$ & $\cdots$ & $\cdots$ & $\cdots$ & $\cdots$ & $\cdots$\\
\hline
\hline
\end{tabular}
\begin{tablenotes}
\item
\textit{Notes.} Column (1): host-galaxy stellar mass. Columns (2)-(8) refer to the (2) $3\sigma$ lower bound; (3) $2\sigma$ lower bound; (4) $1\sigma$ lower bound; (5) median; (6) $1\sigma$ upper bound; (7) $2\sigma$ upper bound; (8) $3\sigma$ upper bound of the sampled $f_\mathrm{occ}$. The table in its entirety is available online.
\end{tablenotes}
\end{threeparttable}
\end{table*}
Due to the fivefold increase in sample size, the findings of this work are significantly more precise than, though still consistent with, previous $f_\mathrm{occ}$ constraints by \citetalias{Miller15} and \citet{Gallo19}, both of which did not find an upper limit for $f_\mathrm{occ}$ in the dwarf galaxy regime. As a consistency check, in Appendix~\ref{append: miller}, we verify that applying the model presented above to fit the same data used in \citetalias{Miller15} does not yield an upper bound on $f_\mathrm{occ}$.\\ 

The key result of this study is that $f_\mathrm{occ}$ decreases sharply with decreasing $M_\star$. The $M_\star$ value at which $f_\mathrm{occ}$ decreases to 50\% is determined to be $\log M_\star = 9.0_{-0.4}^{+0.2}$. \par By convolving the inferred $f_\mathrm{occ}$ with the galaxy stellar mass function derived by \citet{Driver22}, we obtain the following values (1$\sigma$) or lower limits (2$\sigma$):  
\begin{align*}
f_\mathrm{occ}&>93\%&\text{for}&\quad 10^{11} \leq M_\star \leq 10^{12}~M_\odot,  \\
f_\mathrm{occ}&=66_{-7}^{+8}\%&\text{for}&\quad 10^{9} \leq M_\star \leq 10^{10}~M_\odot,  \\
f_\mathrm{occ}&=33_{-9}^{+13}\%&\text{for}&\quad 10^{8} \leq M_\star \leq 10^{9}~M_\odot. 
\end{align*}
The $f_\mathrm{occ}$ uncertainty is asymmetric, and moderate occupation (i.e., $f_\mathrm{occ}\approx 50\%$) is still allowed down to $M_\star=10^8~M_\odot$ at a $2\sigma$ level.\par
%
The median of our sampling results for the probability distribution $p(\lambda)$ of the sARDF is shown as the solid red curve in Figure~\ref{fig: plambda} for galaxies with $M_\star = 10^{10} M_\odot$. The figure also compares the inferred $p(\lambda)$ with results from previous works, which will be discussed in Section~\ref{subsec: plateu}. The main properties of $p(\lambda)$ can be summarized as follows: (i) the distribution exhibits a peak near $\log \lambda \approx 28$; (ii) above this value, $p(\lambda)$ is well approximated by a declining power-law function of $\log \lambda$, up to approximately 31.5, beyond which the bona fide AGN regime begins; (iii) because the best-fit value of $\beta$ in Equation~\ref{eq: mu} is consistent with zero ($\beta = -0.01_{-0.12}^{+0.12}$), $p(\lambda)$ is nearly independent of $M_\star$. 

A peak in the distribution at $\log \lambda \approx 28$ is entirely consistent with the data. Among all the massive galaxies with $M_\star > 10^{10}~M_\odot$ in our sample, 42\% have detected X-ray nuclei with $\log \lambda > 28$ (the actual fraction may be even higher when accounting for undetected sources), implying that $\log \lambda = 28$ approximately corresponds to the median $\lambda$ of massive galaxies.\par
The main results of this work can be intuitively understood as follows: the sARDF and its possible dependence on $M_\star$ are primarily determined by the massive galaxies, which have significantly higher X-ray detection rates compared to low-mass galaxies. The dependence of the sARDF on $M_\star$ is encapsulated by Equation~\ref{eq: mu}, which assumes a linear relationship between $\mu$ (a proxy for $\log \lambda$) and $\log M_{\star}$. This behavior is assumed to hold even in the low-mass regime.
Our sampling yields $\beta \approx 0$, implying that the sARDF of low-mass galaxies is similar to that of massive galaxies. It follows that if most or all low-mass galaxies have central MBHs (i.e., $f_\mathrm{occ} \approx 1$), the model would predict a significantly higher X-ray detection fraction than is measured, even after accounting for the fact that, at a fixed Eddington ratio, low-mass galaxies are less luminous. Our model reflects these processes self-consistently. In Appendix~\ref{append: mock}, we demonstrate that our model can recover $f_\mathrm{occ}$ in both simulated low- and high-$f_\mathrm{occ}$ scenarios.\\

We emphasize that the inferred $p(\lambda)$ is only formally valid within the range $27 \lesssim \log \lambda \lesssim 31$, and extrapolations beyond this range may not be reliable. There is evidence against extrapolating $p(\lambda)$ too far in either direction: (i) at the high $\lambda$ end, it is known that the AGN $p(\lambda)$ follows a broken power-law (e.g., \citealt{Weigel17, Aird18, Zou24b}), and the break is not included in our model; (ii) at the low $\lambda$ end, our $p(\lambda)$ is in tension with observations of the Milky Way MBH Sgr A*, whose X-ray luminosity is $L_\mathrm{X}=2.4\times10^{33}~\mathrm{erg~s^{-1}}$ \citep{Baganoff03}. 
For a host stellar mass of $M_\star = 6.08 \times 10^{10}~M_\odot$ \citep{Licquia15}, this yields $\log \lambda = 22.6$, which is over five orders of magnitude below the inferred peak in $p(\lambda)$. Including the Sgr A* data point in our sample significantly elevates $p(\lambda)$ (compared to the distribution shown in Figure~\ref{fig: plambda}) below $\log \lambda \lesssim 26$, although the range between $27 \lesssim \log \lambda \lesssim 31$ remains virtually unchanged and still exhibits a peak at $\log \lambda \approx 28$. However, we find that with the inclusion of Sgr A*, assuming that $p(\lambda)$ follows a Box-Cox distribution would be falsified (cf. Equation \ref{eq: normtest} and Section \ref{subsec: norm_plambda}). This implies that the Box-Cox distribution family is not sufficiently flexible to model the full $p(\lambda)$ distribution (all the way down to $\log \lambda \lesssim 23$ and up to $\log \lambda \gtrsim 33$). We further note that this limitation does not undermine our $f_\mathrm{occ}$ inference, as it is only sensitive to $p(\lambda)$ in the range $28 \lesssim \log \lambda \lesssim 31$.
\par
We anticipate that the occupation fraction of low-mass galaxies would be higher if the sARDF of low-mass galaxies had a peak at lower $\lambda$ values. To precisely determine where the peak in the sARDF would need to be to achieve full occupation in the dwarf galaxy regime, we consider two mass bins: $\log M_\star = 8 \pm 0.5$ and $9 \pm 0.5$. We fix the sARDF shape parameters $(\sigma, \xi)$ to the values shown in Figure~\ref{fig: sampling} and {\it assume} $f_\mathrm{occ} = 1$. Maximizing the likelihood over $\mu$ (which is a proxy for the distribution peak value) returns $\mu = 26.0$ and $27.4$ for the $\log M_\star = 8 \pm 0.5$ and $9 \pm 0.5$ bins, respectively.
This means that, to reproduce the low X-ray detection rates in the case of full occupation, the peaks of the sARDFs of low-mass galaxies would need to shift towards lower values (specifically by 2.1 and 0.7~dex for the $\log M_\star = 8 \pm 0.5$ and $9 \pm 0.5$ bins, respectively). 
However, we emphasize that there is currently no evidence to suggest that MBHs in low-mass galaxies are more likely to have lower $\lambda$ compared to those in massive galaxies. 
This work provides an observational benchmark for the sARDF of massive galaxies in the low-$\lambda$ regime (see Section~\ref{subsec: plateu}). High-resolution simulations of individual galaxies yield predictions that are broadly consistent with our estimates (e.g., \citealt{Yuan18}). However, observational constraints upon the sARDF for low-mass galaxies in the low-$\lambda$ regime are currently unavailable; additional uncertainties driven by stellar feedback, seed mass, and nuclear star clusters (e.g., \citealt{Partmann25, Petersson25}) make theoretical modeling at the low-mass end particularly challenging.

Indirect clues about the accretion properties of low-mass galaxies come from AGNs. Although simulations still face challenges and lack consensus (e.g., \citealt{Sharma22, Beckmann23}), observational studies show that the AGN sARDF and AGN fraction of low-mass galaxies with $10^8 \lesssim M_\star \lesssim 10^{10}~M_\odot$ and $z \lesssim 0.6$ are similar to those of massive galaxies--if AGNs are identified based on $\lambda$ rather than $L_\mathrm{X}$ \citep{Birchall20, Birchall22, Zou23}. 
While we acknowledge that any statistical inference model involves assumptions, we argue that the model presented in this work is both sufficiently flexible and physically reasonable.\\

\par
Our results are in stark contrast to those obtained by \citetalias{Burke25}, which reports a very high $f_\mathrm{occ}$ (close to 100\% down to the lowest mass bin). We will examine the reasons for this disagreement in detail in Section~\ref{subsec: comp_B25}, and provide independent evidence for a plateau in the sARDF[/sERDF] at low $\lambda$[/$\lambda_{\rm Edd}$] in Section~\ref{subsec: plateu} below.
\section{Discussion}
\label{sec: discussion}
\begin{figure*}
\centering
\includegraphics[width=0.67\hsize]{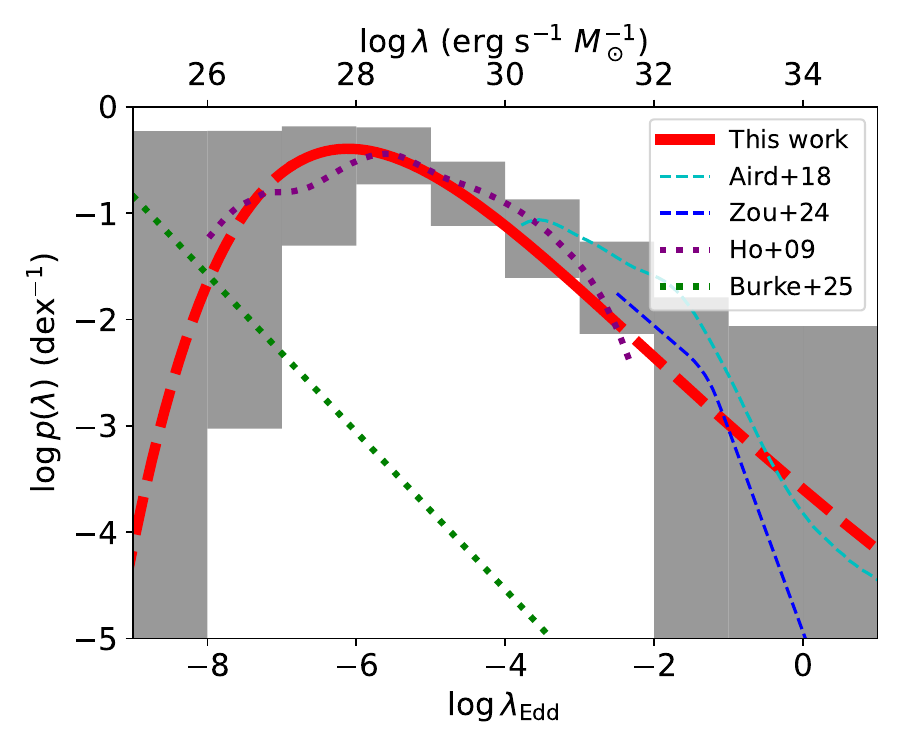}
\caption{The probability distribution, $p(\lambda)$, of the specific accretion rate distribution function for galaxies with $M_\star = 10^{10}~M_\odot$ is shown as a function of $\log\lambda$ (top axis) and $\log \lambda_\mathrm{Edd}=\log\lambda - 34$ (bottom axis). Here, $\lambda$ is converted to $\lambda_\mathrm{Edd}$ by assuming $k_\mathrm{bol} = 25$ and $M_\mathrm{BH} = 0.002M_\star$ (see Section~\ref{subsec: conversion} for a more in-depth discussion). The gray shaded regions bracket the range of allowed $p(\lambda)$ as inferred directly from the data (see Section~\ref{subsec: plateu}), based on a sub-sample of 457 nearby objects with $M_\star > 10^{10}~M_\odot$. The red curve represents our inferred parametric $p(\lambda)$, which is formally valid between $27 \lesssim \log\lambda \lesssim 31$ (shown as a solid line). The probability distributions derived by \citet{Aird18} and \citet{Zou24b} for AGNs are shown as dashed curves for comparison. The probability distributions obtained by \citet{Ho09} and \citetalias{Burke25} based on different samples of nearby galaxies are shown as dotted lines.
}
\label{fig: plambda}
\end{figure*}

This work presents new constraints on the local black-hole occupation fraction, using Chandra imaging data for approximately 1,600 galaxies within a distance of 50 Mpc. We employ a Bayesian model to simultaneously constrain $f_\mathrm{occ}$ and the probability distribution $p(\lambda)$ of the sARDF. We find that $p(\lambda)$ shows little to no evolution with $M_\star$ and exhibits a peak around $10^{28}~\mathrm{erg~s^{-1}}~M_\odot^{-1}$ (Figure~\ref{fig: plambda}). Simultaneously, $f_\mathrm{occ}$ drops below 50\% for host galaxy stellar masses $M_\star \lesssim 10^9~M_\odot$ (Figure~\ref{fig: focc}).
\par 
A key assumption throughout this work is that $p(\lambda)$ follows a Box-Cox distribution (Figure~\ref{fig: illustration}). This distribution is, in principle, more flexible than the normal distribution assumed by \citetalias{Miller15} or the broken power-law distribution (based on \citealt{Weigel17}) assumed by \citetalias{Burke25}, as it can reasonably approximate both functional shapes. In the following sections, we aim to derive model-independent constraints for the sARDF and assess the viability of the aforementioned assumptions, as well as their implications for the inferred occupation fraction, in light of these model-independent constraints. It is important to clarify that $p(\lambda)$ should not be confused with the distribution of bona fide AGNs (i.e., systems with $\log\lambda \gtrsim 32$, which corresponds to $\lambda_\mathrm{Edd} \gtrsim 0.01$), which is indeed well described by a double power law \citep[e.g.,][]{Bongiorno16, Aird18, Yang18, Zou24b}.
Instead, our focus is on constraining the low-luminosity (or quiescent) range below $\log\lambda \lesssim 31$. These systems may exhibit different accretion physics from AGNs--as the accretion flow likely transitions from a geometrically thin disk to an optically thin, radiatively inefficient flow (e.g., \citealt{Yuan14} and references therein). We will further discuss whether the inferred $p(\lambda)$ for low-luminosity galactic nuclei accurately extrapolates into the AGN regime.


\subsection{The Accretion Rate Distribution of Local MBHs}
\label{subsec: comp_sARDF}
\subsubsection{Box-Cox vs. Normal distribution}
\label{subsec: norm_plambda}
First, we verify whether data are statistically consistent with a Box-Cox distribution. By applying the inverse transformation of Equation~\ref{eq: boxcoxtrans}, our assumption becomes equivalent to:
\begin{align}
\frac{\mu}{\xi\sigma}\left[\left(\frac{\log\lambda}{\mu}\right)^\xi-1\right]\sim\mathcal{N}(0, 1),\label{eq: normtest}
\end{align}
We perform a normality test for censored data \citep{Millard13} and obtain a high $p$-value of 0.2, suggesting that a Box-Cox distribution accurately represents $p(\lambda)$.
\par
Next, we evaluate whether a normally distributed $p(\lambda)$, as assumed by \citetalias{Miller15}, is consistent with the data. The Box-Cox distribution simplifies to a normal distribution when $\xi=1$. However, our best-fit value is $\xi=-4.67_{-1.60}^{+1.55}$, which differs from $1$ at a highly statistically significant level. To confirm this discrepancy, we fix $\xi$ at 1 and refit our model. The normality test in Equation~\ref{eq: normtest} yields a low $p$-value of $9\times10^{-4}$, thus refuting the normal distribution assumption. This outcome arises because the data distribution is right-skewed with a heavy tail toward high $\lambda$, characteristics that normal distributions cannot adequately capture.
Moreover, if we adopt the noiseless likelihood in Section~\ref{sebsec: like_noiseless}, the inferred $\log M_\star$:$\log \lambda $ slope under the normal distribution assumption would be $-0.16_{-0.12}^{+0.12}$, corresponding to a $\log M_\star$:$\log L_\mathrm{X}$ slope of $+0.84$. This value is good agreement with the slope reported by \citet{Miller12} and \citet{Miller15}, who argued for downsizing in black-hole accretion--a scenario where lower mass galaxies exhibit relatively higher $\lambda$ values compared to more massive ones. However, after we account for the measurement uncertainties (Section~\ref{subsec: like_final}), the $\log M_\star$:$\log \lambda $ slope becomes $-0.04_{-0.13}^{+0.12}$ under the normal distribution assumption and $-0.01_{-0.12}^{+0.12}$ under our Box-Cox distribution assumption. As Figure~\ref{fig: comp_like} shows, the slope would be slightly underestimated if not accounting for the measurement uncertainties. Therefore, our analysis suggests that previous downsizing results likely no longer hold.\par
\subsubsection{Box-Cox vs. Power-Law Distribution}
\label{subsec: comp_B25}
We specifically address the apparent disagreement with the recent findings of \citetalias{Burke25}, who report a very high occupation fraction across the host stellar mass spectrum. It is important to note that \citetalias{Burke25} use a consistent statistical model to explore various diagnostics of MBH activity, such as nuclear \mbox{X-ray} and radio emissions, as well as optical variability, consistently finding a high occupation fraction across the mass range ($f_\mathrm{occ}>90\%$ at $M_\star \gtrsim 10^8~M_\odot$; at the $2\sigma$ level). In this discussion, we concentrate on the \mbox{X-ray} sample and identify their assumed ERDF as the primary source of the discrepancy.

Although both our study and \citetalias{Burke25} begin with the 50MGC to draw galaxy samples, \citetalias{Burke25} utilizes several \mbox{X-ray} catalogs, including ROSAT, Chandra, XMM-Newton, and eROSITA. In contrast, we restrict our selection to CSC 2.1 sources with ACIS data. An additional difference is the chosen matching radius between CSC 2.1 (or other catalogs) and 50MGC, as well as the aperture over which XRB contamination is evaluated. We emphasize that these differences do not influence the conclusions presented below. In fact, we verified that when we apply a similar Bayesian inference method as \citetalias{Burke25} to the Chandra sample described in Section~\ref{sec: data}, we recover very high occupation fractions down to low galaxy masses, in agreement with \citetalias{Burke25}.
\par
In this work, we assume the same\footnote{Modulo the additional $\delta$ factor in the exponentiation term of Equation~\ref{eq: focc}.} functional expression for the MBH occupation fraction probability as \citetalias{Burke25}, but make different assumptions for the underlying $p(\lambda)$. For clarity, we recall that we make use of the sARDF, which is a function of $\lambda=L_\mathrm{X}/M_\star$, while \citetalias{Burke25} define the ERDF as a function of  $\lambda_\mathrm{Edd}=k_\mathrm{bol}L_\mathrm{X}/L_{\rm Edd}$. A rough conversion between the sARDF and the ERDF is given by $\log\lambda \approx  \log\lambda_\mathrm{Edd}+ 34$ (this is formally correct for an \mbox{X-ray} bolometric correction $k_{\rm bol}=25$ and black-hole to host-galaxy stellar mass ratio $M_\mathrm{BH}/M_\star=0.002$). 
The ERDF assumed by \citetalias{Burke25} is described by a smoothly broken power-law model, whereby the probability of hosting a low-$\lambda_\mathrm{Edd}$ MBH increases monotonically with decreasing $\lambda_\mathrm{Edd}$, down to a minimum value $\log\lambda_\mathrm{Edd,min}$, which is inferred to be $\approx -8.7$ (see their Figure~19). 
This assumption originates from the fact that the probability distribution of bona fide AGN is thought to follow a double power-law (e.g., \citealt{Weigel17,Aird18,Zou24b}). Strong evidence for this comes from the {\it measured} AGN luminosity function \citep{Ueda14}. However, there is no empirical evidence to suggest that the double power-law can be extended to quiescent MBHs, i.e., down to $\log\lambda\approx25$, or $\log\lambda_\mathrm{Edd}\approx -9$.
\par
The assumption of a Box-Cox distribution for the sARDF can indeed model a power-law-like tail across a wide range of $\lambda$ values (Figure~\ref{fig: illustration}), although it implicitly assumes the presence of a peak in the distribution. 
Our prior ($\alpha \sim \mathcal{U}[15, 40]$) allows the peak of $p(\lambda)$ to be as low as approximately 15 (in $\log \lambda$), which is more than 10 orders of magnitude below the inferred peak value of $\log \lambda \approx 28$. This effectively permits an ERDF that is qualitatively similar to that assumed by \citetalias{Burke25}. Appendix \ref{append: mock} directly demonstrates that if the underlying $p(\lambda)$ is indeed a broken power-law, our model should be able to recover it well.
\par
It is also noteworthy that the ERDF in \citetalias{Burke25} is assumed to be independent of black-hole mass. This assumption implies that all galaxies must share the same distribution of $\log L_\mathrm{X} - 1.61 \log M_\star$ (or equivalently, that $\lambda \propto M_\star^{0.61}$), where the 1.61 slope is derived from the chosen \mbh:\mstar\ scaling relation (Equation 3 in \citetalias{Burke25}). In contrast, we do not fix the scaling of $\log M_\star$ with $\log\lambda$; in our model, the slope is a free parameter with a fitted value of $-0.01_{-0.12}^{+0.12}$, compared to the 0.61 assumed by \citetalias{Burke25}. A steeper slope implies that lower-mass galaxies tend to have lower $L_\mathrm{X}$ and are thus less likely to be detected, which in turn favors higher occupation. 
\par
We wish to emphasize that, up to this point, we have not provided compelling reasons to prefer one distribution function over the other. In the following section, we derive \textit{model-independent} constraints on the sARDF over a range of $\lambda$ (or $\lambda_{\mathrm{Edd}}$) that is significantly below the levels typically probed by AGN observations. We also demonstrate the presence of a plateau in the sARDF.

\subsection{Bracketing the Specific Accretion Ratio Distribution Function of Quiescent MBHs}
\label{subsec: plateu}
Robust upper and lower bounds to $p(\lambda)$ can be derived directly from the Chandra data without relying on Bayesian inference. 
To do so, we only consider galaxies with $M_\star>10^{10}~M_\odot$ because their occupation fraction is widely accepted to be 100\%, and, by definition, $p(\lambda)$ can only be defined for sources with central MBHs. There are 457 sources in this mass bin, and 245 (54\%) have compact \mbox{X-ray} nuclei. 
For these galaxies, a meaningful range of $p(\lambda)$ values can be determined by examining two extreme scenarios. Consider a specific accretion ratio bin $\log\lambda$, which includes the values between $\log \lambda_1$ and $\log \lambda_2$. The lowest possible $p(\lambda)$ for that bin occurs when none of the sources with upper limits falls within the range between $\log \lambda_1$ and $\log \lambda_2$. Conversely, the highest possible $p(\lambda)$ occurs when all sources with upper limits higher than $\log \lambda_1$ actually have $\log\lambda$ values falling between $\log \lambda_1$ and $\log \lambda_2$. 
\par
Mathematically, we proceed by labeling $(1, 2, \cdots)$ bins in ascending order of $\log\lambda$ and denote with $N_i^\mathrm{cens}$ the number of censored $\log\lambda$ points within the $i^\mathrm{th}$ bin. The minimum number of sources in the $i^\mathrm{th}$ bin is given by: 
\begin{align}
N_{\mathrm{min},i}=\sum_{k}(1-P_{\mathrm{XRB},k})\mathcal{I}(\log\lambda_k\in\{i^\mathrm{th}~\mathrm{bin}\}),\label{eq: Nmin}
\end{align}
where $k$ runs over all the sources with $M_\star>10^{10}~M_\odot$, 
and $\mathcal{I}(\log\lambda_k\in\{i^\mathrm{th}~\mathrm{bin}\})=1$[/0] if $\log\lambda_k$ falls within [/outside of] the $i^\mathrm{th}$ bin. 
The maximum number of sources is: 
\begin{align}
N_{\mathrm{max},i}=N_{\mathrm{min},i}+\sum_{j\ge i}[N_j^\mathrm{cens}+P_{\mathrm{XRB},k}\mathcal{I}(\log\lambda_k\in\{j^\mathrm{th}~\mathrm{bin}\})].\label{eq: Nmax}
\end{align}
The lower[/upper] bound of $p(\lambda)\Delta\log\lambda$ 
is set to the $2\sigma$ lower[/upper] bound of a binomial proportion confidence interval \citep{Wilson27} with the total number of trials set equal to the number of sources with $M_\star>10^{10}~M_\odot$, and the number of successes being $N_{\mathrm{min},i}$ [/$N_{\mathrm{max},i}$].\par
\par
The shaded gray areas in Figure~\ref{fig: plambda} illustrate the range of allowed sARDFs. Generalizing the above reasoning, $N_{\mathrm{max},i} = 0$ if the bin covers a range that exceeds the highest $\log \lambda$ measured in the sample. $N_{\mathrm{max},i}$ equals the total number of censored data points in the sample if the bin covers a range lower than the smallest $\log \lambda$ in the sample. Note that there are almost no data points, whether censored or uncensored, below $\log\lambda < 26$ and above $\log\lambda > 33$. In these cases, $N_{\mathrm{min},i} = 0$. Furthermore, not all $p(\lambda)$ curves within the viable gray range are permissible, as they must integrate to unity by definition. Critically, this analysis demonstrates that the sARDF ceases increasing with decreasing $\log \lambda$ below $\log\lambda\lesssim28$, or $\log\lambda_{\rm Edd} \lesssim -6$. The inferred sARDF, shown by the red curve in Figure~\ref{fig: plambda}, is comfortably within the range of allowed values. 
\par
Figure~\ref{fig: plambda} also includes a set of published sARDF/ERDF probability distributions, including the ERDF derived by \citetalias{Burke25} (dotted green curve), as well as a representative ERDF based on the results\footnote{\citet{Ho09} assessed the Eddington ratio, $\lambda_\mathrm{Edd}$, for a sample of 175 nearby galaxies with Chandra data but did not explicitly calculate the sample's ERDF. We derive the ERDF based on their dataset. In summary, we first employ the Nelson-Aalen estimator \citep{Aalen78} to compute the cumulative hazard function, $H(\log\lambda_\mathrm{Edd})$, for their $\lambda_\mathrm{Edd}$ distribution. Given that the resultant $H(\log\lambda_\mathrm{Edd})$ is non-differentiable, we apply a spline fit to smooth it and subsequently convert it to $p(\log\lambda_\mathrm{Edd})$ using the relation $p(\log\lambda_\mathrm{Edd})=e^{-H}\frac{dH}{d\log\lambda_\mathrm{Edd}}$. This approach involves significant uncertainties, particularly in the $\log\lambda_\mathrm{Edd}$ intervals that are inadequately represented by the data. Consequently, we consider it only a coarse approximation of their ERDF.} of \citet{Ho09} (dotted purple curve). The \citetalias{Burke25} and \citet{Ho09} ERDFs are shifted by $+33.4$ and $+33.8$, respectively, to allow for a uniform representation in units of $\lambda$.\footnote{The different shifts are applied in order to ensure self-consistency between the definition of $\lambda_\mathrm{Edd}$ in the original works and the definition of $\lambda$ adopted in this work. Specifically, \citetalias{Burke25} adopts $k_\mathrm{bol}=10$ and $M_\mathrm{BH}/M_\star=8\times10^{-5}$ at $M_\star=10^{10}~M_\odot$, whereas \citet{Ho09} adopts $k_\mathrm{bol}=15.8$ and derives the black-hole mass values using the bulge stellar velocity dispersion relation.} For comparison, we also include the sARDFs inferred for AGN using different samples \citep{Aird18, Zou24b}, which are in qualitative agreement with the broken power-law derived by \citet{Weigel17}. \\

A visual comparison of the different probability distributions shows that the sARDF derived in this work is in good qualitative agreement with both the results of \citet{Ho09} (based on a sample with comparable luminosities) and the AGN findings from various groups, whereas it is 
inconsistent with the probability distribution assumed by \citetalias{Burke25}. Aside from the chosen functional shape, the parametrization of \citetalias{Burke25} implies a very low normalization for the ERDF probability distribution above $\log \lambda_{\rm Edd} > \log \lambda_{\rm Edd,min} = -8.7$ (equivalent to $\log \lambda > 24.7$). Above this threshold, the normalization is set by the mass-independent active fraction: $f_a(\lambda_\mathrm{Edd}>\lambda_\mathrm{Edd, min})=\int_{\log\lambda_\mathrm{Edd, min}}^{+\infty}p(\lambda_\mathrm{Edd})\,d\log\lambda_\mathrm{Edd}=0.14 $.  
The resulting ERDF probability distribution, however, is inconsistent with the observed active fraction of massive galaxies. For instance, the bracketing method described in Equations~\ref{eq: Nmin} and \ref{eq: Nmax} yields an active fraction for $\log\lambda>29$ between $13\%-28\%$ for $M_\star>10^{10}~M_\odot$, whereas integrating \citetalias{Burke25}'s ERDF over the same range yields less than 1\%. 
In summary, our comparison with the findings of \citetalias{Burke25} suggests that their chosen ERDF parametrization, combined with the implicit assumption of a steep \lx:\mstar\ relation, leads to the inference of a large population of highly underluminous MBHs.

\subsection{Dependence on Galactic Morphology and Star-forming Properties}
\label{subsec: difftype}
\begin{figure*}
\centering
\includegraphics[width=\hsize]{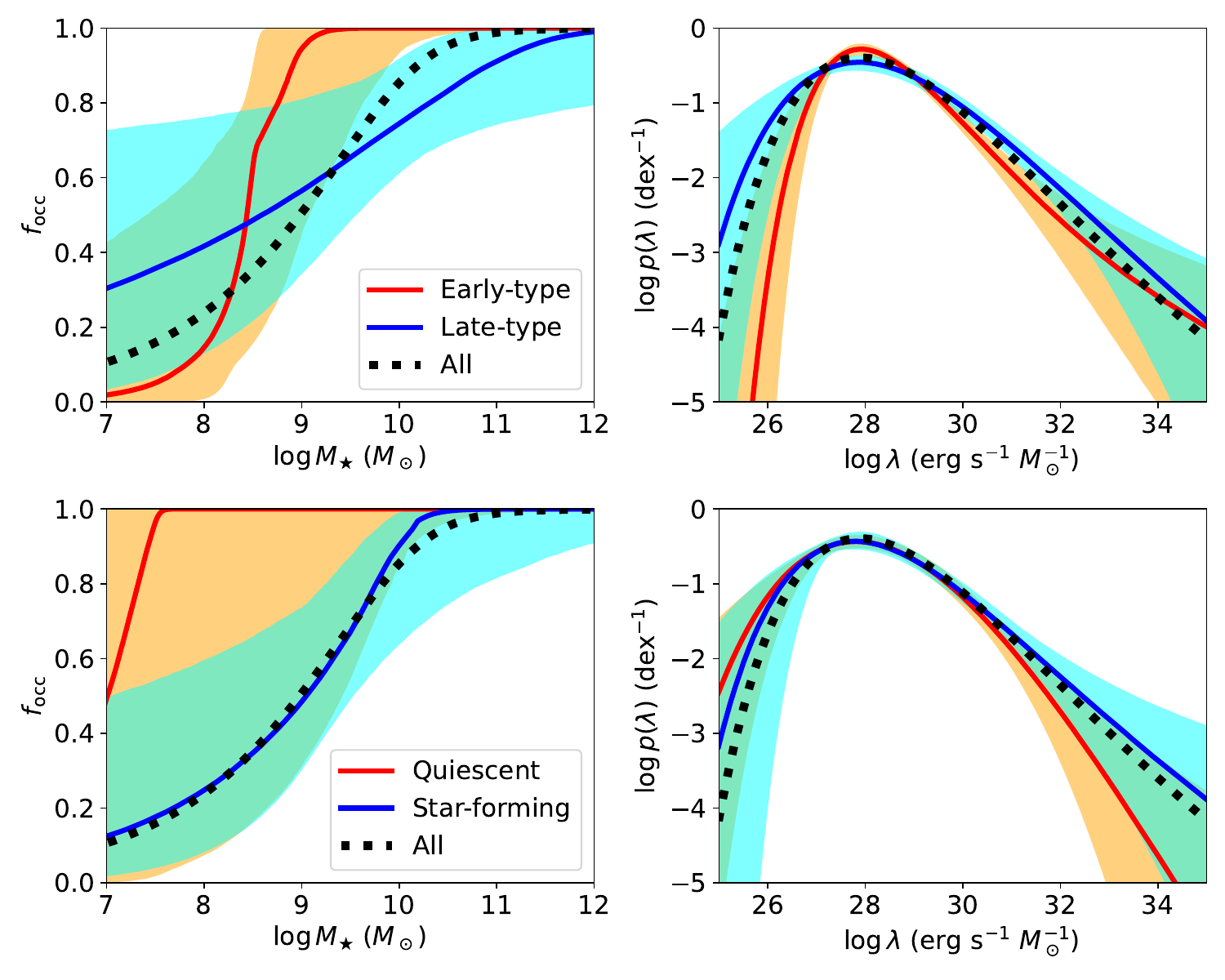}
\caption{Sampled $f_\mathrm{occ}$ (left) and $p(\lambda)$ (right) for different types of galaxies separately. In the top panels, the red and blue solid curves represent the median of the posterior distributions for early-type and late-type galaxies, respectively, while in the bottom panels, they represent quiescent and star-forming galaxies, respectively. The black dotted curves correspond to all galaxies, referenced from Figures~\ref{fig: focc} and \ref{fig: plambda}. The shaded regions indicate the $2\sigma$ uncertainty levels.}
\label{fig: diffgaltype}
\end{figure*}

We further derive $f_\mathrm{occ}$ and $p(\lambda)$ separately for different types of galaxies. Using the 50MGC morphological classifications \citep{Ohlson24}, we divide our sample into 618 early-type galaxies (with 186 compact X-ray detections) and 988 late-type galaxies (with 202 compact X-ray detections). We repeat the analysis as described in Section~\ref{sec: method} for both galaxy types. The results, presented in the top panels of Figure~\ref{fig: diffgaltype}, are qualitatively similar to our original findings, with both early- and late-type galaxies showing a decrease in $f_\mathrm{occ}$ toward lower $M_\star$. Late-type galaxies tentatively exhibit higher $f_\mathrm{occ}$ at $M_\star \lesssim 10^9~M_\odot$ along with elevated $\lambda$, indicating that the corresponding $p(\lambda)$ is higher at $\lambda \gtrsim 10^{29}~\mathrm{erg~s^{-1}}~M_\odot^{-1}$. However, within $2\sigma$ uncertainties, early- and late-type galaxies remain statistically consistent.

Similarly, we examine whether star-forming and quiescent galaxies have similar $f_\mathrm{occ}$ and $p(\lambda)$. A galaxy is classified as quiescent if its SFR is more than one dex below the star-forming main sequence described in \citet{Leja22}; otherwise, it is classified as star-forming. Our sample comprises 625 quiescent galaxies (with 165 compact X-ray detections) and 750 star-forming galaxies (with 150 compact X-ray detections). The corresponding results are shown in the bottom panels of Figure~\ref{fig: diffgaltype}. Star-forming galaxies exhibit a decreasing $f_\mathrm{occ}$ as $M_\star$ decreases, but the constraints for quiescent galaxies are broad and consistent with $f_\mathrm{occ} = 1$. Their $p(\lambda)$ remains largely consistent, although the quiescent $p(\lambda)$ appears suppressed at $\lambda \gtrsim 10^{31}~\mathrm{erg~s^{-1}}~M_\odot^{-1}$. We have verified that the early-type, late-type, quiescent, and star-forming sub-samples with compact nuclear X-ray detections generally have small $P_\mathrm{XRB}$ values (Section~\ref{subsec: xrb}), with median values below 6\%, indicating that XRB contamination is minimal.
\subsection{Eddington Ratio and Mass Accretion Rate Conversions}
\label{subsec: conversion}
Our analysis utilizes $\lambda = L_\mathrm{X} / M_\star$, rather than $\lambda_\mathrm{Edd} = L_\mathrm{bol} / M_\mathrm{BH}$. This approach has the distinct advantage of avoiding the uncertainties associated with converting $(L_\mathrm{X}, M_\star)$ to $(L_\mathrm{bol}, M_\mathrm{BH})$. In principle, the sARDF can be converted to the ERDF using the bolometric correction factor $k_\mathrm{bol} = L_\mathrm{bol} / L_\mathrm{X}$, or even to a mass accretion rate ($\dot{M}$) distribution via the radiative efficiency $\epsilon = L_\mathrm{bol}/(\dot{M}c^2)$. Figure~\ref{fig: plambda} adopts a nominal value of $k_\mathrm{bol} = 25$ when plotting the ERDF, which is typical for standard thin-disk accretion in AGNs. The characteristic AGN radiative efficiency is usually assumed to be $\epsilon = 0.1$ (e.g., \citealt{Davis11, Brandt15}). However, both $k_\mathrm{bol}$ and $\epsilon$ may vary for low-$\lambda_\mathrm{Edd}$ black holes, since the accretion mode is expected to transition to radiatively inefficient, hot accretion flows as $\lambda_\mathrm{Edd}$ decreases (e.g., \citealt{Narayan95, Mahadevan97, Yuan14} and references therein). Importantly, variations in $k_\mathrm{bol}$ and $\epsilon$ do not affect our determination of the sARDF and $f_\mathrm{occ}$, as our results rely exclusively on $L_\mathrm{X}$ rather than $L_\mathrm{bol}$.
\par
With regard to $k_\mathrm{bol}$, \citet{Lopez24} analyze the X-ray-to-infrared SEDs of MBHs with $-8 \lesssim \log\lambda_\mathrm{Edd} \lesssim -3$, similar to the $\lambda_\mathrm{Edd}$ range of our sources (see Figure~\ref{fig: plambda}). Using physical models of advection-dominated accretion flows (ADAFs) and truncated accretion disks, they find that $k_\mathrm{bol}$ converges to a value of 9 in the low-luminosity regime. Although this is lower than the typical AGN $k_\mathrm{bol}$, it remains consistent with extrapolations of the AGN $k_\mathrm{bol}$--$L_\mathrm{bol}$ relation presented in \citet{Duras20}.\footnote{It is worth noting that, for low-$\lambda_\mathrm{Edd}$ MBHs, the ratio of UV/optical luminosity to $L_\mathrm{X}$ deviates significantly from that of AGNs, as radiatively inefficient accretion flows lack the prominent UV/optical emission bump produced by standard thin disks. Nevertheless, $k_\mathrm{bol}$ is less suppressed because the IR emission, where ADAF SEDs peak, contributes substantially to $L_\mathrm{bol}$.} Changing $k_\mathrm{bol}$ from 25 to 9 shifts the ERDF in Figure~\ref{fig: plambda} to the left by approximately 0.4~dex. Alternatively, one can adopt an $L_\mathrm{bol}$-dependent $k_\mathrm{bol}$ (e.g., Equation~4 in \citealt{Lopez24}), which further modifies the shape of the AGN ERDF, since the dependence of $k_\mathrm{bol}$ on $L_\mathrm{bol}$ becomes significant primarily in the AGN regime. Since our sARDF is already determined, converting it to the ERDF using different $k_\mathrm{bol}$ values and the $M_\mathrm{BH}\mathrm{:}M_\star$ relation is straightforward for interested readers. Overall, reasonable variations in $k_\mathrm{bol}$ cause a shift of $\lesssim0.5$~dex, while adopting different $M_\mathrm{BH}\mathrm{:}M_\star$ relations (e.g., \citealt{Shankar20a}) can result in a shift of up to $\lesssim1$~dex.
\par
Regarding $\epsilon$, much larger deviations from the canonical AGN value of 0.1 are expected. We denote $\lambda_M$ as the ratio between $\dot{M}$ and the Eddington mass accretion rate, $\dot{M}_\mathrm{Edd} = 10L_\mathrm{Edd}/c^2$, i.e., $\lambda_M = \dot{M}/\dot{M}_\mathrm{Edd} = (0.1/\epsilon)\lambda_\mathrm{Edd}$. \citet{Xie12} show that $\epsilon$ quickly decreases with decreasing $\lambda_M$ when $\lambda_M \lesssim 0.01$, and the expected $\epsilon$ may be as low as $10^{-4}$–$10^{-3}$ when $\lambda_M = 10^{-6}$. Therefore, $\lambda_M$ is generally larger than $\lambda_\mathrm{Edd}$ by a few orders of magnitude when $\lambda_\mathrm{Edd} \ll 1$. The exact value of $\epsilon$ as a function of $\lambda_M$ depends on the chosen ADAF model parameters, especially $\delta_\mathrm{ADAF}$, the fraction of viscously dissipated energy that heats electrons (see, e.g., \citealt{Yuan14}). Varying $\delta_\mathrm{ADAF}$ within a nominal range of $0.1$–$0.5$ can introduce up to a one-dex uncertainty in $\epsilon$ \citep{Xie12}. Adopting $\epsilon$ as a function of $\lambda_M$ from Table~1 of \citet{Xie12} and $\delta_\mathrm{ADAF}=0.1$ (0.5), we find that $\log\lambda_\mathrm{Edd}=-6$ corresponds to $\log\lambda_M=-4.6\ (-5.2)$ with $\epsilon=0.03\ (0.01)$. Nevertheless, although $\epsilon$ may vary significantly with $\lambda_M$, its changes have no impact on our sARDF or ERDF because $\lambda_\mathrm{Edd}$ is instead based on $k_\mathrm{bol}$, which has a limited variation for low-$\lambda$ MBHs.

\subsection{Implications for the Local Black-Hole Mass Function}
\label{subsec: bhmf}
The distribution of black-hole masses, particularly those in $z=0$ galactic nuclei, is key to understanding MBH growth over cosmic history. For black hole masses below $\approx10^7~M_\odot$, this distribution can elucidate the assembly of MBH progenitors (at $z \simgt 15$; \citealt{Woods19}), affect tidal disruption event rates \citep{Stone16}, and distinguish between models for star-formation suppression in dwarf galaxies \citep{Silk17}. LISA \citep{Amaro-Seoane17} will probe black holes from \mbh\ $= 10^4$ to $10^7$ \msun, though current knowledge is limited \citep{Gallo19, Greene20}.
\par
The black-hole mass function (BHMF), $\phi(M_\mathrm{BH})$, is defined as the number of MBHs per unit comoving volume per unit $\log M_\mathrm{BH}$. The integral of the BHMF over redshift yields the total number of MBHs in the observable universe. In the local universe, MBH masses are typically estimated through empirical scaling relations with a separate observational quantity, such as \mstar, whose distribution is known. Accounting for the \mstar-dependent MBH occupation fraction, the BHMF takes the form:   
\begin{align}
\phi(M_\mathrm{BH})=&\int f_\mathrm{occ}(M_\star)\phi(M_\star)\frac{1}{\sqrt{2\pi\sigma_M^2}}\times\nonumber\\
&\exp\left(-\frac{[\log M_\mathrm{BH}-\mathbb{E}(\log M_\mathrm{BH}|M_\star)]^2}{2\sigma_M^2}\right)d\log M_\star,
\end{align}
where $\phi(M_\star)$ is the galaxy stellar mass function, $\mathbb{E}(\log M_\mathrm{BH}|M_\star)$ is the chosen $M_\mathrm{BH}$:$M_\star$ scaling relation, and $\sigma_M$ is its scatter. 
\begin{figure*}
\includegraphics[width=\hsize]{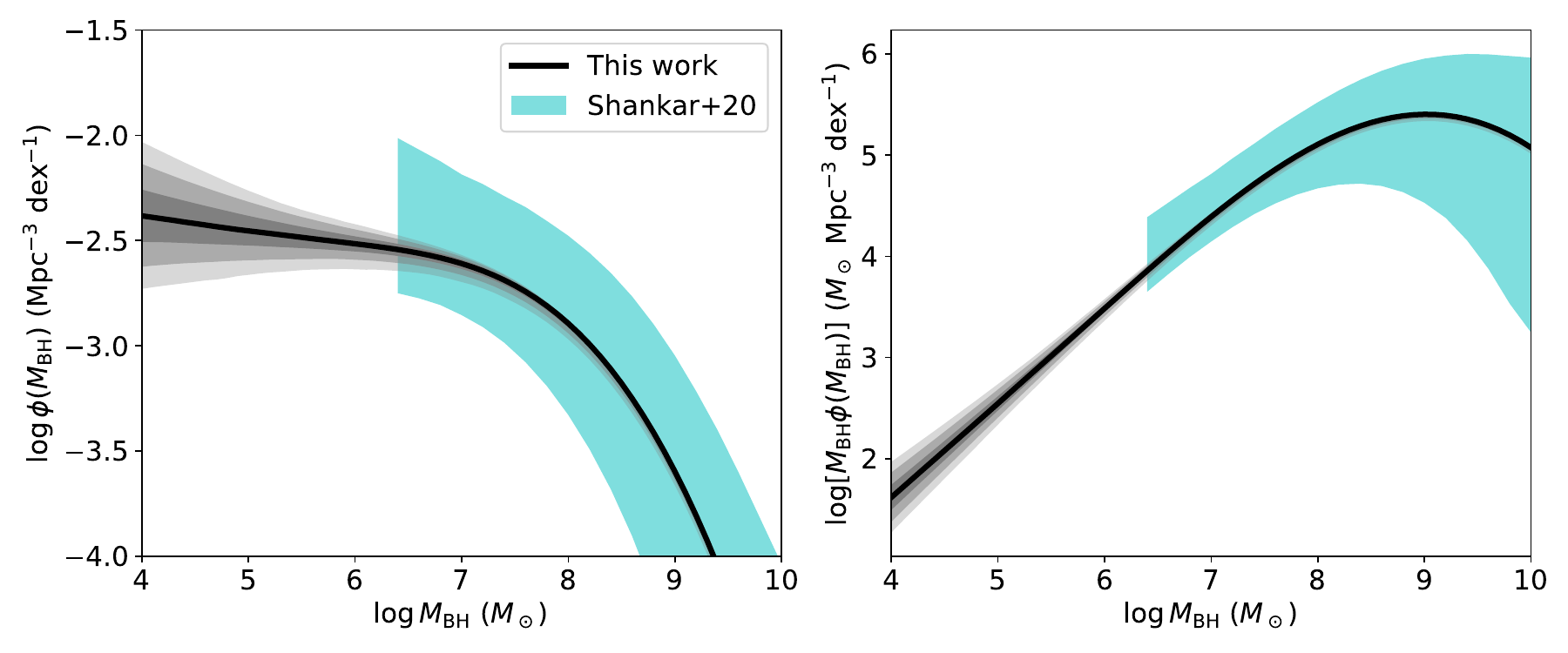}
\caption{The black hole mass function, plotted as $\phi(M_\mathrm{BH})$ and $M_\mathrm{BH} \phi(M_\mathrm{BH})$ in the left and right panels, respectively. The right panel highlights the contributions from MBHs at different masses to the total $M_\mathrm{BH}$ budget. The black curve represents the BHMF based on the occupation fraction derived in this study, with the gray shaded regions indicating the 1, 2, and $3\sigma$ uncertainty ranges, accounting solely for uncertainties in $f_\mathrm{occ}$. The cyan shaded area illustrates the possible BHMF range derived from \citet{Shankar20a}, which is extensive due to the uncertainties in the $M_\mathrm{BH}$:$M_\star$ scaling relation. The effect of a declining $f_{\rm occ}$ is to flatten $\phi(M_\mathrm{BH})$ for $M_\star \lesssim 10^6~M_\odot$.}
\label{fig: bhmf}
\end{figure*}
We adopt the stellar mass function derived by \citet{Driver22} and the same scaling relation as presented in \citet{Gallo19}: $\mathbb{E}(\log M_\mathrm{BH}|M_\star)=8.13+1.72(\log M_\star-11)$ with $\sigma_M=0.61$~dex. This scaling relation is based on the combined AGN and inactive-galaxy sample from \citet{Reines15}. The AGN-based scaling relation and the inactive-galaxy one are found to be different (e.g., \citealt{Reines15}), and we adopt the combined one to mitigate the underlying uncertainties.
The resulting BHMF is shown in Figure~\ref{fig: bhmf}. The gray shaded region represents the statistical uncertainty induced by the $f_\mathrm{occ}$ inference. The effect of a declining occupation fraction flattens the BHMF at $M_\mathrm{BH} \lesssim 10^6~M_\odot$ (e.g., \citealt{Gallo19, Cho24}).
The median BHMF within $10^4<M_\mathrm{BH}<10^{10}~M_\odot$ can be well-approximated by the following polynomial fit:
\begin{align}
\log\phi=&-0.0420-0.941\log M_\mathrm{BH}+0.0592(\log M_\mathrm{BH})^2\nonumber\\
&+0.0142(\log M_\mathrm{BH})^3-0.00156(\log M_\mathrm{BH})^4
\end{align}
The right panel of Figure~\ref{fig: bhmf} illustrates that the total mass budget of MBHs is dominated by supermassive black holes with $M_\mathrm{BH} \gtrsim 10^8~M_\odot$. The contribution from MBHs with $M_\mathrm{BH} \lesssim 10^6~M_\odot$ is expected to be several orders of magnitude lower, regardless of $f_\mathrm{occ}$. We emphasize that the primary uncertainty in constraining the low-mass end of the BHMF arises from the mass scaling relation. The shaded cyan area at high masses in Figure~\ref{fig: bhmf} illustrates the extent of this uncertainty, which exceeds one dex, as noted by \citet{Shankar16, Shankar20a, Shankar20b}.
\par
An important caveat is that the BHMF presented in this work refers only to {\it central} MBHs. It has been argued that a large fraction (potentially over 50\%) of MBHs in dwarf galaxies could be off-nuclear \citep[e.g.,][]{Voggel19, Mezcua20, Greene20, Bellovary21, Sacchi24}. In fact, even in massive galaxies, a significant portion (potentially up to 40\%) of the total $M_\mathrm{BH}$ budget may reside in off-nuclear MBHs \citep[e.g.,][]{Kulier15, Ricarte21b, Pechetti22, Haberle24, Zou24a}. Large volume cosmological simulations also predict the existence of a large population of wandering, off-nuclear MBHs (e.g., \citealt{Ricarte21b, Weller22, DiMatteo23}); but see also \citet{vanDonkelaar25}.
The off-nuclear MBH population remains poorly understood and is therefore not considered in this work. Incorporating an additional model component to account for the off-nuclear fraction will be necessary when connecting our observational results to theoretical models. Directly detecting these off-nuclear MBHs is challenging, as most are dormant and only rarely observed as hyper-luminous X-ray sources, dual AGNs, or off-nucleus tidal disruption events. Nevertheless, some observational constraints are possible, as discussed by, for example, \citet{Guo20} and \citet{Ricarte21a}.

\section{Summary}
\label{sec: summary}
In this work, we present joint observational constraints on the black hole occupation fraction $f_\mathrm{occ}$ of local galaxies and the probability $p(\lambda)$ of the specific accretion rate distribution function (sARDF) of their central MBHs, based on over two decades of Chandra observations of 1,606 nearby galaxies within 50 Mpc. Our main results are summarized as follows:
\begin{enumerate}
\item  $f_\mathrm{occ} \approx 1$ for massive galaxies with $M_\star \gtrsim 10^{10}M_\odot$, but it decreases rapidly at lower $M_\star$. We obtain $f_\mathrm{occ}(10^8 < M_\star < 10^{9}M_\odot) = 33_{-9}^{+13}\%$ and $f_\mathrm{occ}(10^9 < M_\star < 10^{10}M_\odot) = 66_{-7}^{+8}\%$. The $f_\mathrm{occ}(M_\star)$ function is tabulated in Table~\ref{tbl: focc}. See Section~\ref{subsec: sampling}.
\item 
The MBH sARDF, $p(\lambda)$, can be well described by a Box-Cox distribution for inactive galaxies. Its main features include a peak at $\log \lambda \approx 28$ and a power-law tail at $\log \lambda \gtrsim 29$, which can further connect with the AGN $p(\lambda)$ at $\log \lambda \gtrsim 31.5$. $p(\lambda)$ also shows no significant shifts with respect to $M_\star$. We demonstrate that this newly parameterized $p(\lambda)$ is more consistent with the data compared to normal or power-law distributions. See Sections~\ref{subsec: sampling}, \ref{subsec: comp_sARDF}, and \ref{subsec: plateu}.
\item  We examine $f_\mathrm{occ}$ and $p(\lambda)$ separately for different types of galaxies, including morphological classifications (early- vs. late-type) and star-forming properties (quiescent vs. star-forming). We find that the conclusion of decreasing $f_\mathrm{occ}$ at $M_\star \lesssim 10^{10}M_\odot$ also holds for early-type, late-type, and star-forming sub-samples, but we do not detect statistically significant evidence for quiescent galaxies. Nevertheless, these different types of galaxies share statistically consistent $f_\mathrm{occ}$ and $p(\lambda)$, although late-type and star-forming galaxies tentatively have higher $p(\lambda)$ at $\lambda \gtrsim 10^{29}~\mathrm{erg~s^{-1}}~M_\odot^{-1}$. See Section~\ref{subsec: difftype}.
\item We derive the BHMF from the galaxy stellar mass function and the $M_\mathrm{BH}$:$M_\star$ scaling relation, with $f_\mathrm{occ}$ included. The BHMF at $M_\mathrm{BH} \lesssim 10^6M_\odot$ becomes flat. We find that the $f_\mathrm{occ}$ uncertainty is subdominant compared to the scaling-relation uncertainty when determining the BHMF normalization. See Section~\ref{subsec: bhmf}.
\end{enumerate}\par

To the extent that MBH occupation in nearby galaxies serves as a viable diagnostic for MBH seeds at high redshifts (see Section~\ref{sec: intro} and references therein), the results presented in this work appear to favor a heavy seeding scenario. However, further theoretical work is necessary to quantitatively link $f_\mathrm{occ}$ with MBH seeding models. First, as discussed in Section \ref{subsec: bhmf}, there is an additional layer of uncertainty due to the off-nuclear MBH population, which remains poorly understood. Second, the extrapolation of both light-seed and heavy-seed models to low redshifts is subject to significant uncertainties. For example, \citet{Ricarte18b} argues that light-seed models may result in a low $f_\mathrm{occ}$ if most seeds undergo little growth and thus remain undetectable at $z=0$. Conversely, \citet{Bhowmick24} contends that heavy-seed models could yield a high $f_\mathrm{occ}$ if MBH seed formation is more prevalent. Despite these challenges, this work provides a crucial observational benchmark for constraining this fundamental quantity, and future research can further refine MBH seeding models based on $f_\mathrm{occ}$.

\begin{acknowledgments}
We thank the anonymous referee for constructive comments and suggestions. We are grateful to Riccardo Arcodia and Feng Yuan for helpful insights. This research has made use of data obtained from the Chandra Source Catalog, provided by the Chandra X-ray Center (CXC). Support for this work was provided by the National Aeronautics and Space Administration through Chandra Award Number 21700098 issued by the Chandra X-ray Center, which is operated by the Smithsonian Astrophysical Observatory for and on behalf of the National Aeronautics Space Administration under contract NAS8-03060. A.C.S acknowledges support from National Science Foundation astronomy and astrophysics grant AST-2108180. W.N.B. acknowledges support from NSF grant AST-2407089. A.E.R. gratefully acknowledges support provided by NSF through CAREER award 2235277.
\end{acknowledgments}

\appendix
\section{X-ray-detected sources with $M_\star < 10^8~M_\odot$}
\label{append: lowmass}

Three sources with $M_\star < 10^8~M_\odot$ are detected by Chandra. Such cases, where possible MBHs reside in very low-mass galaxies, are often interesting individually, and there may not be a consensus regarding their nature. We do not aim to incorporate their complexities in this work and only briefly comment on their properties as noted in the literature.

\begin{itemize}
    \item PGC~027864: R.A. = 09:44:01.87, decl. = $-00:38:32.1$, distance = 21~Mpc, $\log M_\star = 7.4$, and $\log L_\mathrm{X}=39.0$. This galaxy is metal-poor and discussed in detail by \citet{Reefe23}. It contains another even brighter X-ray source $3''$ away \citep{Senchyna20} that is not included in our analyses. The galaxy nucleus has a broad H$\alpha$ line, implying an estimated $M_\mathrm{BH} \approx 3150~M_\odot$. \citet{Reefe23} showed that the galaxy nucleus has an [\iona{Fe}{x}]~$\lambda6374$ coronal line, which is strong evidence indicating AGN activity; however, this line may also be a misidentified \iona{Si}{ii}~$\lambda6371$ line \citep{Herenz23}.

    \item PGC~031259 (also known as Mrk~1434): R.A. = 10:34:10.15, decl. = +58:03:49.1, distance = 36~Mpc, $\log M_\star = 7.1$, and $\log L_\mathrm{X} = 39.9$. This galaxy is metal-poor and discussed specifically by \citet{Thygesen23}. It contains two luminous X-ray sources that can only be explained by either ULXs or AGNs. Its high $L_\mathrm{X}$ also leads to its independent selection as a dwarf AGN by XMM-Newton \citep{Birchall20, Cann24} and eROSITA \citep{Bykov24}.

    \item PGC 1206166: R.A. = 15:08:22.69, decl. = +01:47:55.0, distance = 24~Mpc, $\log M_\star = 8.0$, and $\log L_\mathrm{X} = 38.8$. This galaxy is an early-type member of the NGC~5846 group \citep{Marino16}. Its X-ray emission was reported by \citet{Miller12} and is unlikely to be explained by XRBs.
\end{itemize}

\section{Selection Effects}
\label{append: selection}
Two main selection effects are likely to affect our sample. First, the parent galaxy catalog, the 50MGC, is incomplete for dwarf galaxies.  Second, Chandra ACIS may be biased in favor of accretion-powered MBHs. For the first selection effect, we find that the galaxies considered in this work become systematically more compact than the expected mass-size relation in \citet{Carlsten21} at $M_\star\lesssim10^8~M_\odot$, indicating that we may be missing low surface brightness systems. We expect that this effect has only a limited systematic impact on our inference of $f_\mathrm{occ}$. To examine this, we consider the sub-sample of galaxies within 30~Mpc and with $M_\star > 10^8~M_\odot$; as shown in Section~4.2 of \citet{Ohlson24}, galaxies in this regime should be complete. The resulting ACIS-covered sample size is reduced to 1,007 galaxies (63\% of the original sample), with 277 compact \mbox{X-ray} detections. We rerun our analyses for this mass-complete subsample and verify that the median $f_\mathrm{occ}$ is similar to that of the original sample, though with somewhat larger uncertainties (not explicitly plotted here).\par
The second selection effect likely has a more direct impact, as the galaxies in our sample were initially proposed for Chandra observations for a variety of scientific reasons, a fraction of which could be related to the presence of MBHs. The Chandra proposal subject categories include \texttt{``solar system and exoplanets''}, \texttt{``stars and white dwarfs''}, \texttt{``white-dwarf binaries and cataclysmic variables''}, \texttt{``black-hole and neutron-star binaries''}, \texttt{``supernova, supernova remnant and isolated neutron star''}, \texttt{``gravitational wave event''}, \texttt{``normal galaxies''}, \texttt{``active galaxies and quasars''}, \texttt{``clusters of galaxies''}, \texttt{``extragalactic diffuse emission and surveys''}, and \texttt{``galactic diffuse emission and surveys''}. Among these, only \texttt{``active galaxies and quasars''}, and in some cases \texttt{``normal galaxies''}, are directly relevant to central MBHs; the other categories are generally unrelated.
Therefore, for simplicity, we assume that the Chandra selection effects are primarily driven by targeted observations of individual sources under the \texttt{``active galaxies and quasars''} and (potentially) \texttt{``normal galaxies''} categories. To account for this, we attempt to remove all sources within one arcminute of the centers of these specific observations. If we exclude sources only from the \texttt{``active galaxies and quasars''} category, 1,312 sources (82\% of the original sample size) remain in our sample, with 263 compact detections. If we instead exclude sources from both the \texttt{``active galaxies and quasars''} and \texttt{``normal galaxies''} categories, 847 sources (53\% of the original sample size) remain, with 120 compact detections.
We note that this removal procedure may introduce an opposite selection bias, namely, that high-$\lambda$ sources could become underrepresented in the final sample.

\begin{figure*}
\includegraphics[width=\hsize]{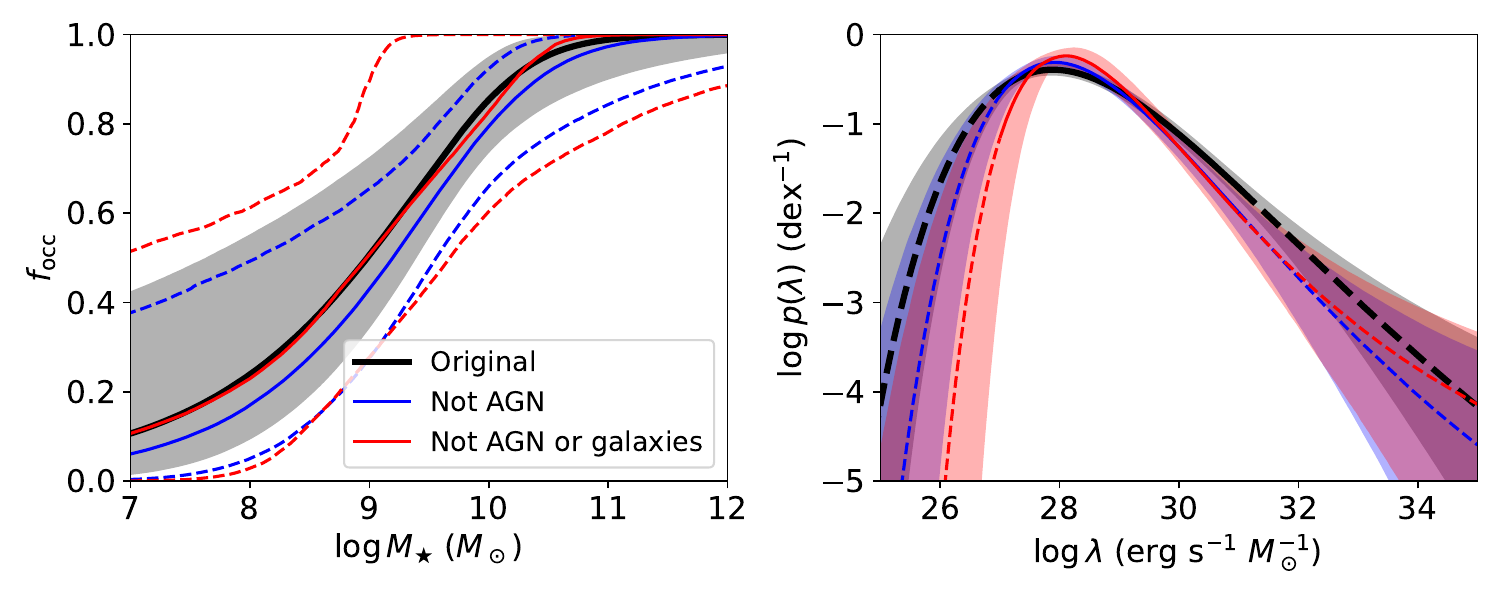}
\caption{Comparison of $f_\mathrm{occ}$ (left) and $p(\lambda)$ (right). Black curves represent the original results. Blue curves correspond to the sample excluding all sources within one arcminute of observational centers with the Chandra proposal subject category \texttt{``active galaxies and quasars''}; red curves indicate the sample further excluding sources based on both the \texttt{``active galaxies and quasars''} and \texttt{``normal galaxies''} categories. In the left panel, solid curves show the median of the posterior distributions, while the shaded regions and dashed curves indicate the $2\sigma$ ranges. In the right panel, curves represent the medians, with dashed segments showing extrapolations beyond the data range (as in Figure~\ref{fig: plambda}), and shaded regions correspond to the $2\sigma$ ranges.}
\label{fig: serendipitous}
\end{figure*}

We rerun our analysis and present the corresponding results in Figure~\ref{fig: serendipitous}. Compared to the original results, the median $f_\mathrm{occ}$ values are similar, and $p(\lambda)$ is only slightly suppressed at the high-$\lambda$ end. Overall, the measurements remain largely consistent, indicating that the selection effects associated with intentional Chandra observations of MBH-related phenomena are smaller than the statistical uncertainties.
However, it is unlikely that selection effects can be fully eliminated. While they may be mitigated by using samples from dedicated surveys of specific sky regions (e.g., the Virgo Cluster; \citealt{Gallo10}), the resulting smaller sample size, as demonstrated by \citetalias{Miller15}, would lack sufficient constraining power.

\section{Results Based on M15}
\label{append: miller}

\begin{figure}
\includegraphics[width=\hsize]{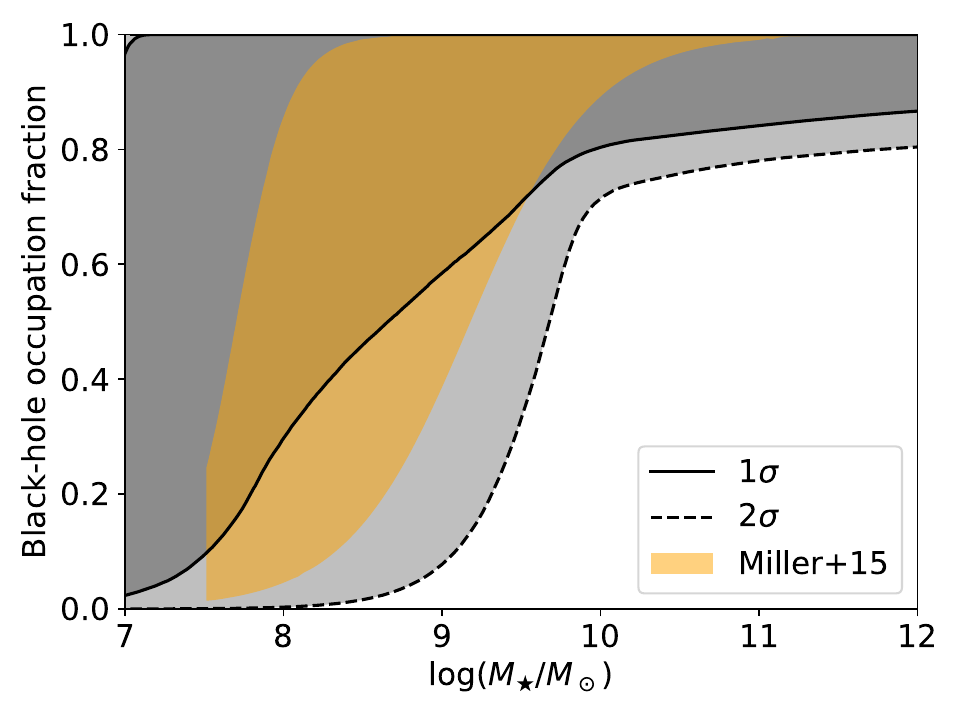}
\caption{$f_\mathrm{occ}$ results when applying our methodology to the \citetalias{Miller15} data. The solid and dashed curves, along with their corresponding shaded regions, represent the $1\sigma$ and $2\sigma$ ranges, respectively. The original $1\sigma$ measurements from \citetalias{Miller15} are shown as the orange region for comparison. In agreement with \citetalias{Miller15}, our results show no evidence of decreasing $f_\mathrm{occ}$ at low $M_\star$.}
\label{fig: focc_miller}
\end{figure}

We apply the methodology presented in the main text to the data in \citetalias{Miller15} to examine if we can obtain consistent results. Their sample included approximately 200 optically selected early-type galaxies within 30 Mpc, which is a subsample of ours. We adopt their ``clean'' sample from Table~1 and present the $f_\mathrm{occ}$ results in Figure~\ref{fig: focc_miller}. Both our measured $f_\mathrm{occ}$ and the original \citetalias{Miller15} $f_\mathrm{occ}$ are consistent with $f_\mathrm{occ} = 1$; the \citetalias{Miller15} $f_\mathrm{occ}$ constraint appears narrower outside the range $10^8 \lesssim M_\star \lesssim 10^{10}~M_\odot$ because \citetalias{Miller15} assumed $f_\mathrm{occ}(M_\star < 10^7~M_\odot) \approx 0$ and $f_\mathrm{occ}(M_\star > 10^{10}~M_\odot) \approx 1$. The overall consistency serves as further evidence that the low $f_\mathrm{occ}$ inferred in the main text is indeed genuine based on improved data--if we downgrade the data quality, we would not detect the decreasing trend of $f_\mathrm{occ}$ at $M_\star < 10^{10}~M_\odot$.

\section{Inference for Mock Data}
\label{append: mock}

To validate our methodology, we generate mock data and examine our fitting results. Section 2.2 of \citetalias{Burke25} conducted similar simulations and validated the overall consistency when recovering $f_\mathrm{occ}$. In this Appendix, we primarily focus on examining whether we can still reasonably recover $p(\lambda)$ and $f_\mathrm{occ}$ if the low-$\lambda$ $p(\lambda)$ instead follows a power-law. We extrapolate the sARDF at $M_\star = 10^{10}~M_\odot$ and $z = 0.05$ from \citet{Zou24b} to low-$\lambda$ using a power-law. Since the integral of $p(\lambda)$ is one by definition, the distribution must have a sharp decrease below a minimum $\log \lambda$, which we infer to be 28.4. We set $f_\mathrm{occ}(M_\star \ge 10^{9.5}~M_\odot)=1$ and consider two estreme cases for dwarf galaxies: $f_\mathrm{occ}(M_\star < 10^{9.5}~M_\odot)=0$ or 1. For each $M_\star$ value in our sample, we randomly assign a $\lambda$ value as per Equation~\ref{eq: base}. We then adopt an $L_\mathrm{X}$ threshold of $10^{39.4}~\mathrm{erg~s^{-1}}$, ensuring that the fraction of massive galaxies with $M_\star > 10^{10}~M_\odot$ above the threshold matches the fraction of uncensored data points at $M_\star > 10^{10}~M_\odot$ in our sample (54\%). All sources with $L_\mathrm{X}$ below this threshold are treated as upper limits (note that the minimum $\log \lambda$ and the $L_\mathrm{X}$ threshold are higher than those in the main text because a power-law extrapolation of $p(\lambda)$ itself is inconsistent with observations, as discussed in Section~\ref{subsec: plateu}). 

\begin{figure*}
\includegraphics[width=\hsize]{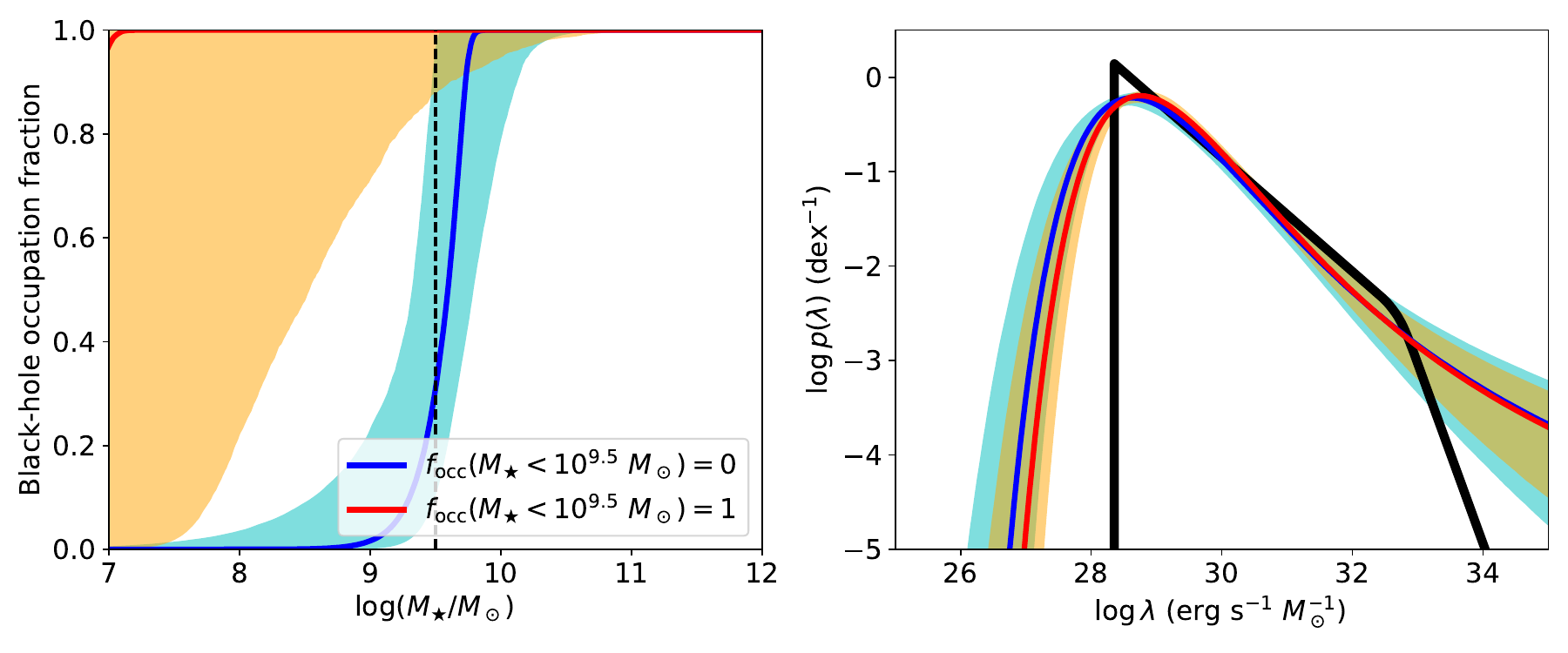}
\caption{The recovered $f_\mathrm{occ}$ (left) and $p(\lambda)$ (right) from mock data generated based on the $p(\lambda)$ shown as the black curve in the right panel. Our inputs are $f_\mathrm{occ}(M_\star \ge 10^{9.5}~M_\odot)=1$ and $f_\mathrm{occ}(M_\star < 10^{9.5}~M_\odot)=0$ (blue) or 1 (red). The blue and red solid curves represent the medians of the fitted posteriors, and the shaded regions indicate the $2\sigma$ levels. The vertical dashed line in the left panel marks $M_\star = 10^{9.5}~M_\odot$. The fitted results are largely consistent with the inputs.}
\label{fig: simu}
\end{figure*}

We use the same method as in Section~\ref{sec: method} to measure $p(\lambda)$ and $f_\mathrm{occ}$ for the mock data, and the results are presented in Figure~\ref{fig: simu}. The fitted $f_\mathrm{occ}$ values are in good agreement with the input $f_\mathrm{occ}$ in both low- and high-$f_\mathrm{occ}$ cases. For $p(\lambda)$, our fitted curves capture the general trend well; the peak is slightly larger than the minimum $\lambda$, and the high-$\lambda$ tail broadly resembles the power-law tail. Small differences are present because our Box-Cox distribution family does not exactly replicate a power-law distribution with a sharp drop.

\bibliography{citations}{}
\bibliographystyle{aasjournalv7}
\end{document}